\documentclass[usenatbib,usegraphicx]{mn2e}
\usepackage{aas_macros}
\usepackage{natbib}
\usepackage{amsmath,amssymb,amsfonts,bm}
\usepackage{subfigure}
\usepackage{longtable}
\usepackage{graphicx}
\usepackage{graphics}
\usepackage[dvipsnames]{color}
\usepackage{keyval}
\usepackage{trig}
\usepackage{float}
\usepackage{calc}
\usepackage{hyperref}

\def\spose#1{\hbox to 0pt{#1\hss}}
\def\lta{\mathrel{\spose{\lower 3pt\hbox{$\mathchar"218$}}
    \raise 2.0pt\hbox{$\mathchar"13C$}}}
\def\gta{\mathrel{\spose{\lower 3pt\hbox{$\mathchar"218$}}
    \raise 2.0pt\hbox{$\mathchar"13E$}}}
\def\beq{\begin{equation}}
\def\eeq{\end{equation}}
\def\bey{\begin{eqnarray}}
\def\eey{\end{eqnarray}}
\def\LOS{{\small LOS}}
\def\ETG{{\small ETG}}
\def\ETGs{{\small ETG}s}
\def\RMS{{\small RMS}}

\def\re{\, {\rm R_e} }

\def\kpc{\, {\rm kpc} }
\def\kpch{\, {\rm kpc~h^{-1}} }
\def\msun{M_\odot}

\def\Msun{M_\odot}
\def\Msunh{\rm M_\odot~h^{-1}}

\def\kms{\, {\rm km \, s}^{-1} }

\def\a0{$a_0$}

\begin{document}
\title{The mass and angular momentum distribution of simulated massive
early-type galaxies to large radii}

\author[Wu et al.]{Xufen Wu$^{1,3}$, Ortwin Gerhard$^{1}$, Thorsten Naab$^{2}$,
Ludwig Oser$^{2}$, Inma \and
Martinez-Valpuesta$^{1}$, Michael Hilz$^{2}$, Eugene Churazov$^{2,4}$,
Natalya Lyskova$^{2,4}$\\
$^{1}$Max-Planck-Institut f\"{u}r Extraterrestrische Physik, Giessenbachstrasse, 85748 Garching, Germany\\
$^{2}$Max-Planck-Institut f\"{u}r Astrophysik, Karl-Schwarzschild-Str. 1, 85741 Garching, Germany\\
$^{3}$Argelander-Institut f\"{u}r Astronomie der Bonn Universit\"{a}t, Auf dem H\"{u}gel 71, 53121 Bonn, Germany\\
$^{4}$Space Research Institute (IKI), Profsoyuznaya 84/32, Moscow 117997, Russia}

\date{Accepted 2013 December 11.  Received 2013 December 6; in original form 2012 September 16}
\pagerange{\pageref{firstpage}--\pageref{lastpage}}
\pubyear{2013}

\maketitle

\begin{abstract}

  We study the dark and luminous mass distributions, circular velocity
  curves (CVC), line-of-sight kinematics, and angular momenta for a
  sample of 42 cosmological zoom simulations of galaxies with stellar
  masses from $2.0\times 10^{10}\Msunh$ to $3.4\times
    10^{11}\Msunh$. Using a temporal smoothing technique, we are able
    to reach large radii. We find that:
    (i) The dark matter halo density profiles outside a few kpc follow
    simple power-law models, with flat dark matter CVCs for lower-mass
    systems, and rising CVCs for high-mass haloes.  The projected
    stellar density distributions at large radii can be fitted by
    S\'ersic functions with $n\gta 10$, larger
      than for typical early-type galaxies (\ETGs).
  (ii) The massive systems have nearly flat total (luminous plus dark
  matter) CVCs at large radii, while the less massive systems have
  mildly decreasing CVCs. The slope of the circular velocity at large
  radii correlates with circular velocity itself.
  (iii) The dark matter fractions within the projected stellar
    half mass radius $\re$ are in the range $15$-$30\%$ and
    increase to $40$-$65\%$ at $5\re$. Larger and more massive
    galaxies have higher dark matter fractions. The fractions and
    trends with mass and size are in agreement with observational
    estimates, even though the stellar-to-total
    mass ratio is  $\sim$2-3 times higher than estimated for \ETGs.
  (iv) The short axes of simulated galaxies and their host dark matter
  haloes are well aligned and their short-to-long axis ratios are
  correlated.
  (v) The stellar root mean square velocity $v_{\rm rms}(R)$ profiles
  are slowly declining, in agreement with planetary nebulae
  observations in the outer haloes of most \ETGs.
%
  (vi) The line-of-sight velocity fields ${\bar v}$ show that rotation
  properties at small and large radii are correlated. Most radial
  profiles for the cumulative specific angular momentum parameter
  $\lambda(R)$ are nearly flat or slightly rising, with values in
  $[0.06,~0.75]$ from $2\re$ to $5\re$.  A few cases show local maxima
  in $|{\bar v}|/\sigma (R)$. These properties agree with observations
  of \ETGs\ at large radii.
  (vii) Stellar mass, ellipticity at large radii $\epsilon (5\re)$,
  and $\lambda (5\re)$ are correlated: the more massive systems have
  less angular momentum and are rounder, as for observed \ETGs.
  (viii) More massive galaxies with a large fraction of accreted stars
  have radially anisotropic velocity distributions outside $\re$.
  Tangential anisotropy is seen only for galaxies with high fraction of
  in-situ stars.

\end{abstract}



\section {Introduction}

Recent observations and cosmological simulations suggests a two-phase formation scenario for massive early-type galaxies (ETGs), in which an early phase of rapid star formation driven by cold accretion and dissipative mergers is followed by a prolonged phase of mass accretion through gas-poor major and minor mergers. Observations have shown that a population of old, massive ($M_{\star}\sim 10^{11}\msun$) and red ETGs were already in place at redshifts $z=2-3$ \citep[e.g.,][]{Fontana_etal2006, Ilbert_etal2010, Cassata_etal2011}, and that these galaxies have sizes several times smaller and densities an order of magnitude higher than present-day ETGs of similar mass \citep[e.g.,][]{Daddi_etal2005,Trujillo_etal2007,vanDokkum_etal2010}. Recent simulations have found that massive galaxies grow initially through rapid star formation fuelled by infall of cold gas at $z \gtrsim 2$, leading to an old population of 'in
situ' stars. Subsequently, the simulated galaxies grow through minor
mergers, accreting old stars formed in subunits outside the main
galaxy halo. The accreted stars are preferentially added to the outer
haloes of the host systems, leading to efficient size evolution
\citep[e.g.,][]{Naab_etal2009, Oser_etal2010, Feldmann_etal2010,
  Hopkins_etal2010, Johansson_etal2012}.

While there is relatively little stellar mass in the outer regions of
ETGs at galactocentric radii $\gtrsim 2 \re$, these stars may carry
indispensable information about the late assembly history of the
galaxies. With relaxation times in the present-day outer stellar haloes
of up to several Gyrs, the record of the recent halo accretion history
may be relatively well preserved \citep{vanDokkum2005,Duc_etal2011}. The outer halo stars may also contain a significant
fraction of the angular momentum of the galaxies
\citep{Romanowsky+Fall_2012}, and they can be used as gravitational
tracers to study the mass distribution, dark matter fraction and
potential of \ETGs\ at large radii. Therefore, a detailed
investigation of the structure and kinematics of the outer stellar
haloes, comparing simulated and observed haloes, may reveal important
information about the formation history of \ETGs.

Traditional long slit measurements and more recent observations with
integral-field units (IFUs) have provided detailed kinematic and
dynamical information about the central regions of \ETGs, $R \lesssim
1-2\re$ \citep[e.g.,][]{Bender_etal1994, Gerhard_etal2001,Cappellari_etal2006}.  Using 2-dimensional stellar kinematics from
the SAURON IFU out to $\sim 1 \re$, and a measure of the projected
angular momentum condensed into the cumulative $\lambda(R)$ parameter
(see \S\ref{sec-lambda}), ETGs can be separated
into two main groups, fast rotators (FR, $\lambda(R) \gtrsim 0.1$) and
slow rotators \citep[SR, $\lambda(R) \lesssim
0.1$][]{Emsellem_etal2007,Emsellem_etal2011}. This division is part of
a wider dichotomy between oblate-spheroidal disky, coreless, rotating
ETGs with little radio and X-ray emission, and triaxial boxy, cored,
non-rotating, radio-loud and X-ray bright systems
\citep[see][]{Bender_etal1989, Kormendy_Bender1996,Kormendy_etal2009}.

In the outer haloes, obtaining kinematic information is much harder
because of the rapid decline of stellar surface brightness with
radius.  Most of the known kinematic properties come from observations
of Planetary Nebulae (PNe) which have been found to be good tracers of
the stars and can be observed up to $\sim 8 \re$
\citep[e.g.,][]{Mendez_etal2001, Coccato_etal2009,McNeil-Moylan_etal2012}. Recently, individual IFU pointings and
slitlet masks have also been used at $3-4\re$
\citep{Weijmans_etal2009, Proctor_etal2009, Murphy_etal2011}. The PNe
observations show that the division between FR and SR is largely
preserved to large radii, although there are also galaxies whose
$\lambda(R)$-profiles drop significantly outwards, possibly implying
that the high inner values are due to disks whose light contribution
fades towards large radii. Most of the PNe \RMS\ velocity profiles
decrease slowly with radius, but a subset of galaxies show steep
'quasi-Keplerian' outer decreases of velocity dispersion
\citep{Coccato_etal2009}.

Analysis of outer stellar kinematics, strong lensing, and hydrostatic
equilibrium of X-ray emitting hot gas shows that massive elliptical
galaxies have nearly isothermal inner mass distributions, equivalent
to flat circular velocity curves \citep[e.g.,][]{Gerhard_etal2001,Koopmans_etal2006, Auger_etal2010, Churazov_etal2008,Churazov_etal2010, Nagino_etal2009}. For the lower mass ellipticals,
the situation is less clear, as the mass-anisotropy degeneracy is
stronger for declining velocity dispersion profiles, their X-ray
emission is too faint, and the lensing samples are dominated by
massive systems. The dynamical modelling of integrated light and PNe
indicates somewhat more diffuse dark matter haloes in these galaxies
\citep{ngc3379, Napolitano_etal2009, Morganti_etal2013}. Another
useful tracer of the outer mass distributions is the globular clusters
(GCs); especially massive ellipticals contain large GC populations
\citep[e.g.,][]{Schuberth_etal2010, Strader_etal2011}. However, a
larger fraction of GCs may be recently accreted systems, as their
relative frequency is tilted more towards small systems than that of
PNe and light \citep{Coccato_etal2012}.

In order to compare this large body of work with cosmological
predictions, we here present the first detailed analysis of the inner
and outer dynamics of a large sample of simulated galaxies. We study
the mass distributions, outer kinematics, and angular momentum
distributions of 42 resimulated galaxies from a high-resolution
cosmological simulation, which grew through the two-phase processes of
early in-situ formation followed by late accretion and minor mergers
\citep{Oser_etal2012}. In \S\ref{ics} we briefly describe the
present-day (z=0) model galaxies extracted from the cosmological zoom
simulations, and the method we use to derive smooth kinematic maps
from their particle distributions. In \S\ref{mass}, we study the mass
density distributions of the stars and dark matter in these galaxies,
as well as the corresponding circular velocity curves (CVCs). Then in
\S\ref{kin}, we investigate the observable kinematics in these systems
out to large radii. We finally consider the cumulative and local
angular momentum profiles $\lambda(R)$ and $|{\bar v}|/\sigma (R)$ for
the stellar components in \S\ref{rotator}. We end by summarizing our
results in \S\ref{conc}.

\section{The simulated galaxies and their kinematic analysis}\label{ics}

The ``galaxies'' studied here are extracted from the cosmological zoom
simulations of \citet{Oser_etal2010,Oser_etal2012}. These simulations
were carried out with the following cosmological model parameters (in
standard notation): $h=0.72,~\Omega_b=0.044,~\Omega_{\rm DM}=0.216,
\sigma_8=0.77$, and initial slope of power spectrum $n_s=0.95$. First,
dark matter-only initial conditions were evolved from $z=43$ to $z=0$,
and selected individual haloes were identified together with their
virial radii $r_{\rm vir}\equiv r_{200}$. For this simulation, the
softening radius was $2.52~\kpch$.  Then these haloes were traced
back in time, and were replaced with high resolution gas and dark
matter particles. The new haloes were evolved again from $z=43$
to the present day including prescriptions for star formation,
supernova feedback, gas cooling and a redshift dependent UV background
radiation.

From these high-resolution simulations, we here select 42 galaxies at
z=0 which do not have massive satellites at this time. The selection
is based on the circular velocity curves of these systems within 5
effective radii ($5~\re$); we require that the estimated fluctuations
in the CVC induced by satellites are smaller than $2\%$.  The final sample of simulated galaxies includes 32 galaxies described in \citet{Oser_etal2010}, and another 10 less massive galaxies from the same cosmological simulation to extend the mass range of the simulated galaxies, which then ranges from $2.0\times10^{10}\Msunh$ to $3.4\times 10^{11}~\Msunh$ within $10\%~r_{\rm vir}$. Their typical effective radii are $\re\simeq$~$1-5~\kpch$ \citep{Oser_etal2010}. The effective radius is here defined as the projected half mass radius of the stellar particles within $10\%~r_{\rm vir}$.  For comparison, the co-moving softening lengths for stars and dark matter particles in the resimulations are $0.4~\kpch$ and $0.89~\kpch$, so the stellar kinematics are resolved for $R>0.5\re$ in the smallest systems, and for $R> 0.2\re$ in the large galaxies.  Individual values of $R_e$ are given for all sample galaxies in column 1 of Table \ref{fitting}.

The star formation model used for the simulations presented here (see
\citealt{Oser_etal2010} for all details) favours efficient star formation
at high redshift leading to preferentially spheroidal systems with old
stellar populations. The simulations do not produce supernova driven
winds and a model for feedback from central AGN is not
included. Therefore, the fraction of available baryons (in every halo)
converted into stars of the central galaxies in the simulated mass
range is typically two to three times higher than estimates from
models matching observed galaxy mass functions to simulated halo mass
functions \citep[e.g.,][]{Guo_etal2010, Moster_etal2010,Behroozi_etal2010, Yang_etal2012}. Possible physical processes
responsible for this discrepancy are strong wind-driving feedback from
SNII \citep[e.g.,][]{Dekel+Silk_1986, Oppenheimer+Dave_2008,
Governato_etal2010, DallaVecchia+Schaye_2012} and/or feedback from
super-massive black holes \citep[e.g.,][]{Croton_etal2006,
DiMatteo_etal2008, McCarthy_etal2010, Teyssier_etal2011}.

Other simulations with similar specifications (weak supernova
feedback and no AGN feedback) result in galaxies with photometric
and kinematic properties similar to present day elliptical galaxies
\citep{Naab_etal2007, Johansson_etal2009, Naab_etal2009,
Feldmann_etal2010, Johansson_etal2012}. The galaxies used here are
in agreement with early-type scaling relations of mass with radius
and stellar velocity dispersion. In addition, they have close to
isothermal total mass distributions, similar to some observed
ellipticals (e.g. \citealp{Gerhard_etal2001, Koopmans_etal2006,
Churazov_etal2010, Barnabe_etal2011}), and their observed size
evolution between $z \sim 2$ and $z = 0$ is in agreement with recent
observational estimates \citep{Oser_etal2012}.

In contrast to the central regions of the simulated galaxies which are
well resolved, in their outer regions ($R > 2\re$) the particle noise can
be substantial.  In addition, although we have already removed systems
having massive satellites, many smaller substructures are still
present in the outer parts of the remaining galaxies. We have decided
to smooth out these small substructures rather than taking them out
one by one. However, we have tested in a few cases that the results
with both approaches are very similar. To reduce fluctuations in the
final velocity fields caused by either particle noise or small
satellites, we use an N-body code \citep[NMAGIC, implemented by][]{nmagic}
to temporally smooth the system while integrating the
orbits of the particles in the gravitational potential, for one
circular orbit period at $10\re~(\sim10\%~r_{\rm vir})$. Since the
total mass in gas within $10\%~r_{\rm vir}$ in these systems is small
(it is one order of magnitude smaller than the mass of stars within
$10\%~r_{\rm vir}$ for all 42 galaxies), we simply fix the gas
particles at their initial positions while integrating the stellar and
dark matter particle orbits.

The NMAGIC code is a spherical harmonics code with a made-to-measure
(M2M) algorithm \citep{Syer_tremaine1996,nmagic}. Here we use it as
a normal N-body code. The Poisson solver adopted in NMAGIC is a
spherical harmonic expansion potential solver. It uses a radial grid
binned logarithmically within maximum radius $r_{\rm max}=360\kpch$,
and there are 400 radial bins. The spherical harmonics expansion is
carried to $l_{\rm max}=16$. This does not resolve the small
satellites, which are therefore conveniently smoothed out during the
integration.

The potential and gravitational acceleration are computed at a
sequence of time steps separated by 
\beq \label{dts} 
{\rm dts}= {1\over 1000} {2\pi \times10\re\over v_{\rm circ,10\re}},
\eeq 
i.e., $1000$ times in one circular orbit period for $r=10\re$. During
each interval {${\rm dts}$} the orbits of the particles are integrated
with an adaptive leap-frog scheme. Kinematic line-of-sight (\LOS)
observables are projected on a two-dimensional polar grid with
resolution $n_r\times n_\phi=20\times 30$ on a $(10\re)^2$ region,
using eq.~(11) of \citet{ngc4697}.  The
observables, e.g., surface density or velocity moments, 
here denoted by $\Delta_j$ for the $j_{\rm th}$ grid cell, are then time
averaged by integrating \citep{Syer_tremaine1996,nmagic}
\beq\label{smooth} 
{\tilde \Delta}_j(t)=\alpha \int_0^\infty 
\Delta_j (t-\tau) e^{-\alpha \tau} d \tau, 
\eeq 
where $\Delta_j(t-\tau)$ is the corresponding
observable for the $j_{\rm th}$ grid cell and the snapshot at time
$t-\tau$, and the temporal smoothing parameter $\alpha$ is taken to
be \beq \alpha \equiv \frac{5}{1000 \, {{\rm dts}}}, \eeq so that the
smoothing time is ${\alpha}^{-1}=200\,{{\rm dts}}$. This procedure
effectively smoothes over the particle noise for the observables in
the sparse outer regions of the simulated galaxies. In a later section
(\S \ref{kin}), we will illustrate the effect of the particle noise
and of the satellites on the projected kinematics for one snapshot and
compare with the time averaged kinematics obtained by the procedure
just described.

\section{Mass distributions and Circular Velocity Curves}\label{mass}\

In this section, we investigate the stellar and dark matter density
profiles of the simulated galaxies at z=0. We use the amplitude and
slope of the circular velocity curve to characterize the mass
distribution, and determine the fraction of dark matter at
intermediate radii.

\begin{figure}{}
\begin{center}
\resizebox{8.cm}{!}{\includegraphics{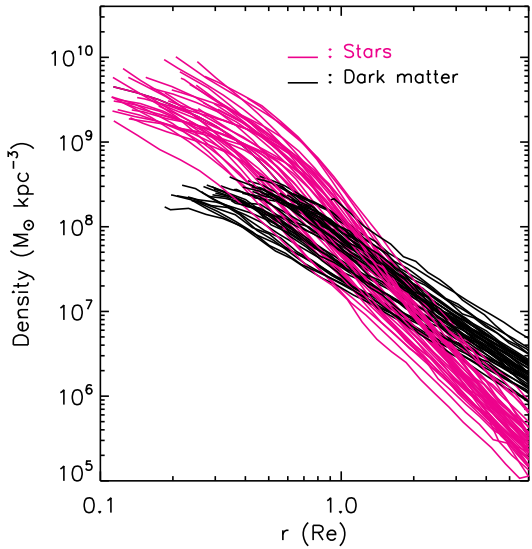}}
\caption{Temporally smoothed mass density profiles for stellar
components (magenta lines) and dark matter haloes (black) of the
simulated galaxies. The profiles are shown as a function of
normalised $r/\re$, extending from the softening length ($0.4
\kpch$ for stars and $0.89\kpch$ for dark matter particles) to
$6\re$.  }\label{density}
\end{center}
\end{figure}

Figure~\ref{density} shows the three-dimensional volume density
profiles for stars and dark matter, for radii greater than the
respective softening radii for the star and dark matter particles.
The densities shown are temporally smoothed as described above. We
find for all model galaxies that the stellar components have steeply
decreasing density profiles from $1\re$ to $6\re$, with small cores in
the centres, while the dark matter haloes have flatter density
profiles.  Since the high resolution re-simulations of individual
haloes include a variety of physical processes for the baryonic
component, the dark matter density profiles are different from the
simple NFW-like profiles \citep{NFW1996, Navarro_etal2010} found in
simulations that only include dark matter. Compared
with the density profiles obtained from an equivalent dark
matter-only cosmological simulation \citep{Oser_etal2010}, the halo
density profiles from the simulation with baryons investigated here
are more cuspy within $1\re$ of their stellar components. The
evolution of the dark matter profiles is dominated by adiabatic
contraction but is also complicated by expansion of the inner halo
due to minor mergers \citep{Johansson_etal2009}.  A detailed study
of the combined processes is beyond the scope of this paper. 

\subsection{Stellar density profiles}\label{stellarfit}
The surface density profiles of observed \ETGs\ can be well fitted
with a S\'ersic profile \citep{Sersic1963,Capaccioli1989}. Luminous
\ETGs\ have `cored' profiles for which the slope of the central
profile is below that of the outer S\'ersic profile
\citep{Trujillo_etal2004}.  We therefore use a cored S\'ersic
profile \citep{Graham_etal2003} to represent the stellar density of
the model galaxies, such that 
\bey\label{sersic}
  I(R)&=& I_0 \exp \left\{ -b_n(n)
   \left[ \left({R'\over R'_e} \right)^{1/n} -1\right] \right\},\\
  R'&=&\sqrt{R^2+R_0^2},\;\;\;\; b_n(n)=1.9992n-0.3271.
\eey
Here $n$ is the S{\'e}rsic index, $I_0$ is a normalization parameter
related to the central surface density, and $R'_e$ is a radius close
to the effective radius if $R_0/R'_e\ll 1$ (because the relation used
for $b_n$ versus $n$ is for the original S\'ersic profile).  We fit
the model density profile out to a truncation radius of $10\%~r_{\rm vir} \sim 10\re$, binning the temporally smoothed surface densities
on a logarithmic radial grid (see \S \ref{ics}).

\begin{figure}{}
\begin{center}
\resizebox{9.cm}{!}{\includegraphics{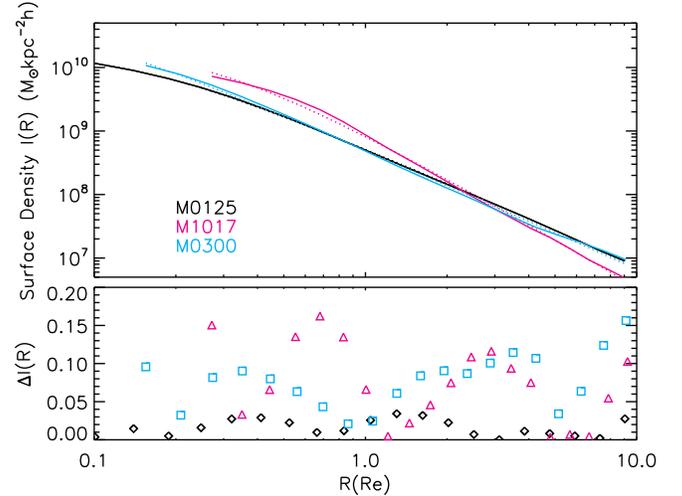}}
\caption{ S\'ersic fits for stellar density profiles. Upper
panel: temporally smoothed stellar surface density profiles I(R)
for three simulated galaxies (solid lines, different colours),
together with cored S\'ersic fits (dotted lines).  Lower panel:
residuals $\Delta I \equiv \frac{|I_{\rm fit}-I_{\rm smooth}|}{I_{\rm smooth}}$ at each fitting point.
}\label{densityfit}
\end{center}
\end{figure}

\begin{figure*}{}
\begin{center}
\resizebox{8.cm}{!}{\includegraphics{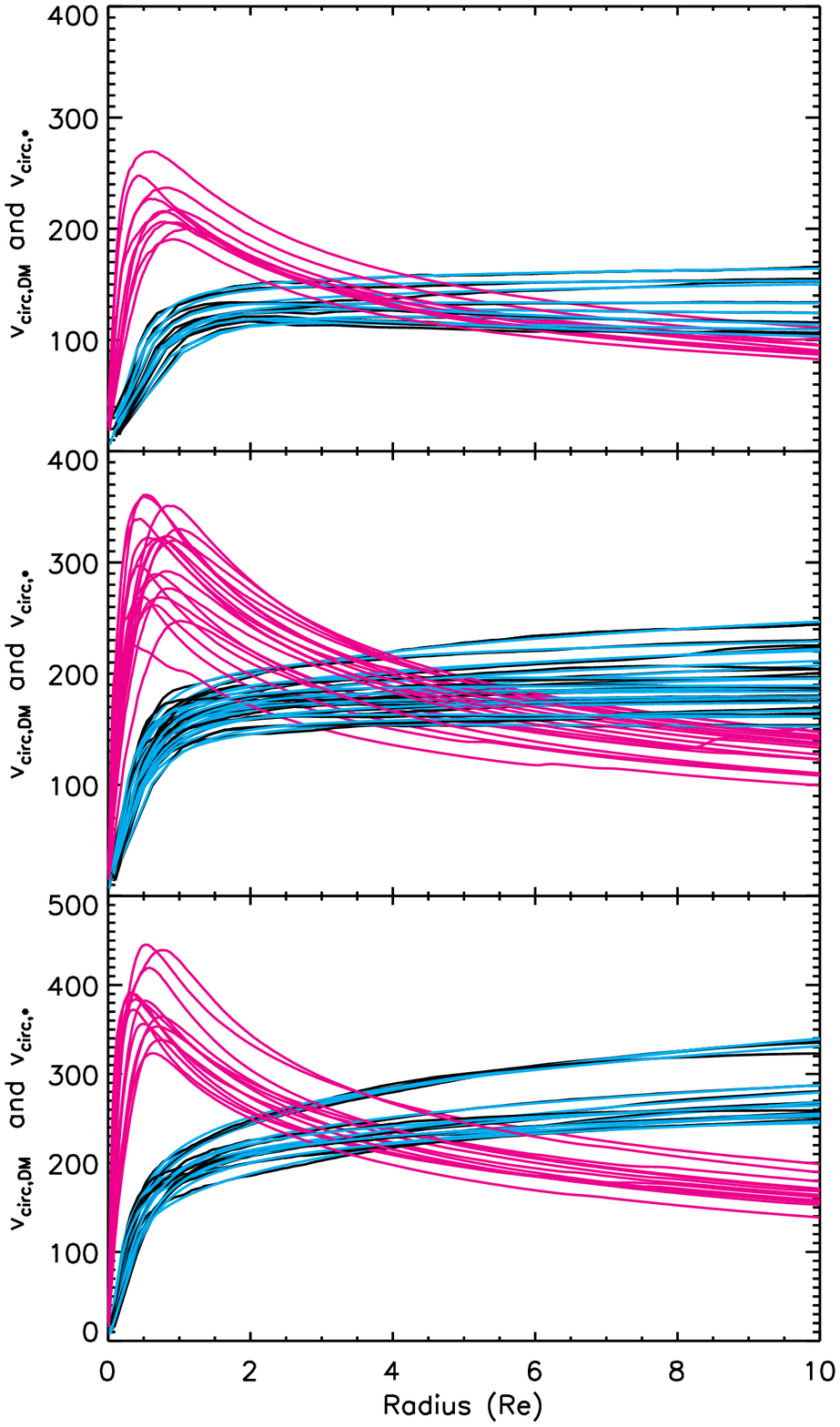}}\resizebox{8.cm}{!}
                   {\includegraphics{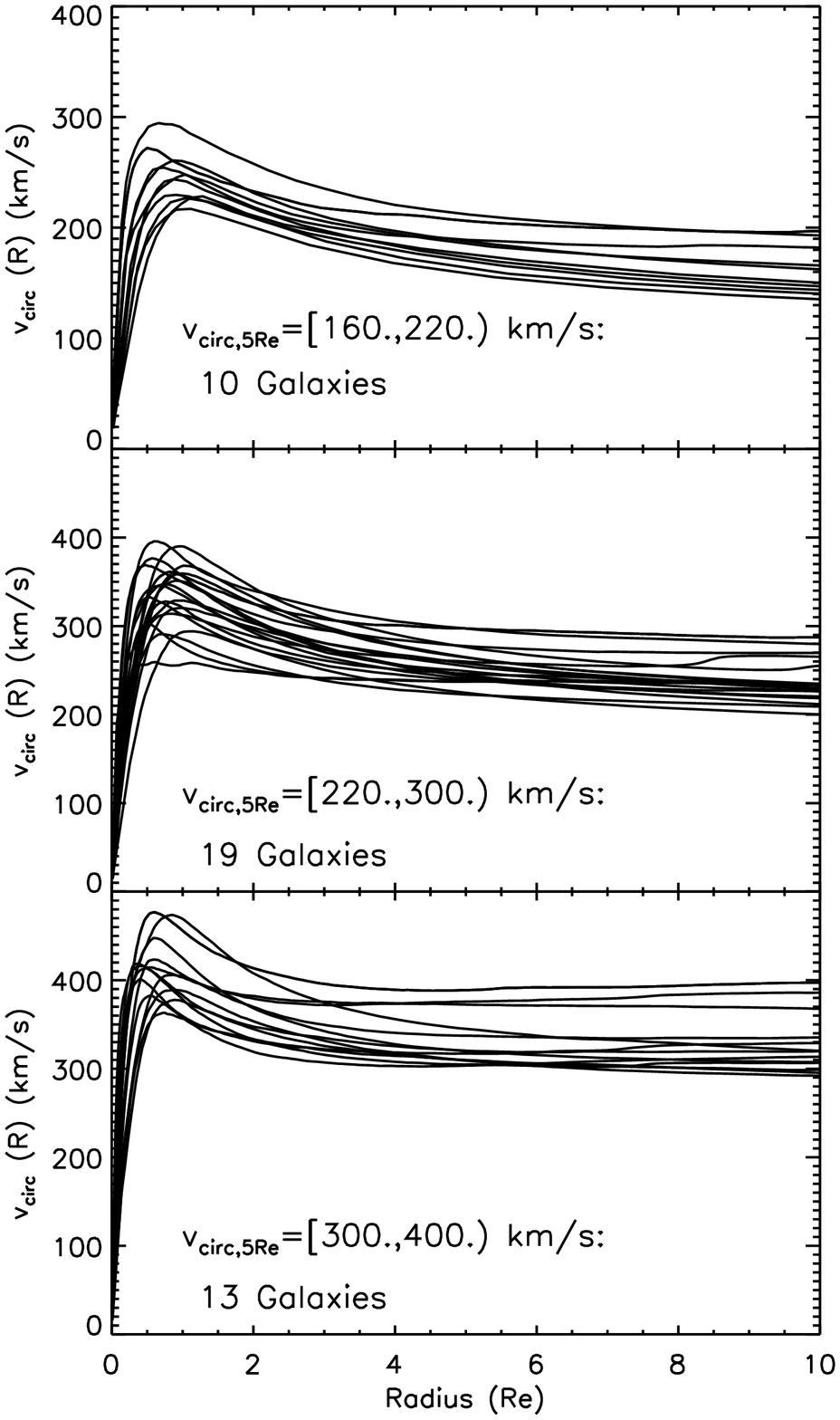}}
\caption{{Left panels:} Contribution to circular velocity curves
(CVCs) from stars (magenta) and dark matter (black) for all model
galaxies, separated in bins of increasing total circular velocity at
$5\re$, as given on the corresponding right panels. The cyan lines
show the parameterised fits to the dark matter CVCs, using
eq.~\ref{vceq}. Massive haloes with high $v_{\rm circ}$ have
increasing dark matter CVCs (bottom panel), while low-mass haloes
have flat CVCs (top). {Right panels:} Total CVCs for all model
galaxies: massive galaxies have flat total CVCs at large radii
(bottom panel), while less massive galaxies (top) have mildly
decreasing CVCs.} \label{vc_vcbin}
\end{center}
\end{figure*}

Figure~\ref{densityfit} shows surface density profiles, cored
S{\'e}rsic fits, and residuals for three model galaxies (M0125,
M1017, and M0300, selected based on their rotational properties, see
\S~\ref{kin} and \ref{rotator}).  Generally, the cored-S\'ersic
model fits the particle distributions well, with a typical residual
(standard deviation) for the entire sample of $10\%$ of the local
surface density. However, most of the model galaxies have large
values of S{\'e}rsic index $n>10$, and a significant fraction even
$n\gg10$, so that these galaxies have more power-law like surface
density profiles than real ETGs.

\subsection{Dark matter mass distributions and circular velocity curves}

The density profiles for the simulated dark matter haloes, shown in
Figure~\ref{density}, are very similar to each other, with nearly the
same slope between $1\re$ and $6\re$.  To describe the mass
distributions of the simulated galaxies more quantitatively, we study
their circular velocity curves (CVCs) in Fig.~\ref{vc_vcbin}.  Here
the circular velocity $v_{\rm circ}(r)=[GM(r)/r]^{1/2}$ serves as a
measure of the spherical part of the mass distribution, independent of
the actual rotational velocities of the stellar or dark matter
particles. The three rows of Fig.~\ref{vc_vcbin} show the model galaxy
CVCs binned into three groups according to the value of $v_{\rm circ}$
at $5\re$. The left panels show the CVCs for the
stellar components (magenta) and dark matter haloes (black)
separately. We see that the dark matter CVCs are nearly flat at large
radii for the lowest $v_{\rm circ}$ bin, but increase outwards for
higher mass systems (with high $v_{\rm circ}$).  The figure also shows
that the dark matter haloes dominate the CVCs outside $4\re$.
The dark matter CVCs can be well approximated by power law mass
distributions between $(1-2)\re$ and $10\re$. We use the
following parametrisation:
\beq \label{vceq} v_{\rm circ}^{\rm DM}=
\frac{v_0}{(5\re)^a} \frac{r^{1.0+a}}{\sqrt{r_c^2+r^2}},
\eeq Here $v_0$ is a normalization parameter for the circular
velocity, $r_c$ is the core radius of the dark halo, and $a$ is the
slope of the CVC at large radii.  The cyan lines in the left panels of Fig.~\ref{vc_vcbin} show the parameterised CVCs of the dark halo components. They agree well with the binned (unsmoothed) data for the halo circular velocities, especially at large radii (black lines).  Columns 7-9 in Table~\ref{tabvc} give the
best-fitting values of $v_0$, $a$ and $r_c$ for the model dark matter
haloes. Column 10 gives the rms residuals of the fit, which have magnitudes of only a few km/s.  The values of $r_c$ range between
$0.72-1.44\kpch$ for almost all models, less than about twice the
softening radius of the dark matter particles, $r_s^{\rm DM}=0.89~\kpch$, and the $r_c$ do not correlate with the masses of
the model galaxies.  Therefore the presence of a core could be an
effect of the softening in the simulations. In order to check this
further, we also redid the fits while fixing $r_c=r_s^{\rm DM}$.  The
best-fitting CVCs obtained in this case are not significantly
different from the previous CVCs (the typical residual is less than
$1\%$). Since $r_c/(5\re) \ll 1$ (Table~\ref{fitting}), the core
radii do not significantly affect the rotation curves around
$5\re$. Thus $v_0$ and $a$ are approximately equal to the value and
logarithmic slope of the dark matter CVC at $5\re$, 
\beq 
v_0\simeq
v_{\rm circ}^{\rm DM}(5\re), \qquad a\simeq S_{5}^{\rm
 DM}\equiv\frac{d\ln v_{\rm circ}^{\rm DM} }{d\ln r}(5\re).  
\eeq

The dark matter CVCs are well-described by power-laws in the
  range $1$-$10 R_e$ for the massive galaxies, and in the range
  $2$-$10 R_e$ for the lower-mass systems; note the small residuals of
  the CVC fits ($\sim 2\kms \sim 1\%$). The range of slopes from -0.1
  to +0.3 corresponds to density slopes of -2.2 to -1.4 down to
  several kpc, indicating that the inner dark matter density profiles
  have been modified by the interaction with the baryonic component.

\begin{figure}{}
\begin{center}
\resizebox{9.0cm}{!}{\includegraphics{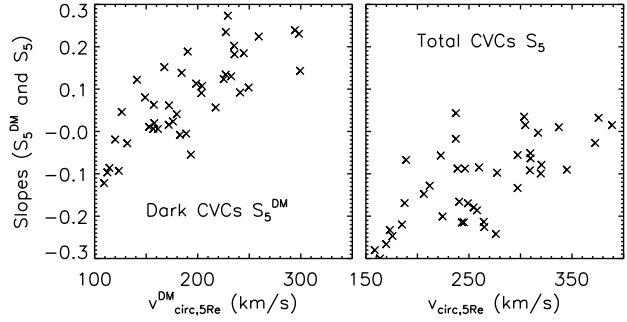}}
\caption{Left panel: The logarithmic slope of the dark matter CVCs for all model galaxies, versus their outer circular velocity, $v_{\rm circ}^{\rm DM}(5\re)$.  Right panel: The logarithmic slope of the total CVCs including stars and gas, versus total $v_{\rm  circ}(5\re)$. Massive galaxies (with large values of $v_{\rm circ}(5\re)$) have zero slopes, i.e., flat CVCs, while smaller galaxies have CVC slopes between flat (0.0) and slightly falling (-0.3).}
\label{vc_slope}
\end{center}
\end{figure}

The left panel of Fig.~\ref{vc_slope} quantifies the correlation
between the outer slope $S_{5}^{\rm DM}$ of the dark matter CVC
with the amplitude of the dark matter CVC at $5\re$ for the simulated
galaxies. We find that for the low-$v_{\rm circ}^{\rm DM} (5\re)$
systems the slopes of dark matter CVCs are around zero and for the
high-$v_{\rm circ}^{\rm DM} (5\re)$ systems these slopes are slightly
above zero ($<0.3$), confirming the result from Fig.~\ref{vc_vcbin}.

The right panels of Fig.~\ref{vc_vcbin} show the {\sl total} CVCs of
the model galaxies, including the contribution from the stars, which
we parametrize again by their value and logarithmic slope at $5\re$,
\beq 
v_{\rm circ}(5\re), \qquad S_{5}\equiv\frac{d\ln v_{\rm  circ}}{d\ln r}(5\re).  \eeq 
The total CVCs are slightly falling at
large radii ($5\re$) for the systems in the upper right panel, whose
$v_{\rm circ}(5\re)$ is smaller than $220\kms$, while they are nearly
flat outside $2\re$ for the most massive model galaxies whose $v_{\rm  circ}(5\re)$ is larger than $300\kms$ (lower right panel; see also
\citealt{Lyskova_etal2012}). Thus the massive model galaxies represent well massive ETGs whose CVCs are nearly isothermal as inferred from dynamical and lensing studies \citep[e.g.,][]{Gerhard_etal2001, Koopmans_etal2006, Auger_etal2010,Churazov_etal2008, Churazov_etal2010, Nagino_etal2009}.

The right panel of Fig.~\ref{vc_slope} quantifies this correlation by
showing the total slopes $S_{5}$ versus the circular velocities at
$5\re$. Model galaxies with larger values of $v_{\rm circ}(5\re)$ have
$S_{5}\sim 0.0$, i.e., flat CVCs at large radii, while the remaining
galaxies have slopes between $[0.0,~-0.3]$, i.e., CVCs between flat
(0.0) and mildly falling (-0.3).

Thus it is clear that the slope of the CVC at large radii is
correlated with the amplitude of the circular velocity, and hence with
model galaxy mass. The correlation is clearest for the dark matter
haloes alone (see left panel of Fig.~\ref{vc_slope} and
Eq. ~\ref{vceq}), while it is weakened when the baryonic component is
taken into account. All values for circular velocities and CVC slopes
at $5\re$ can be found in Table \ref{tabvc}.

\begin{figure}{}
\begin{center}
\resizebox{8.5cm}{!}{\includegraphics{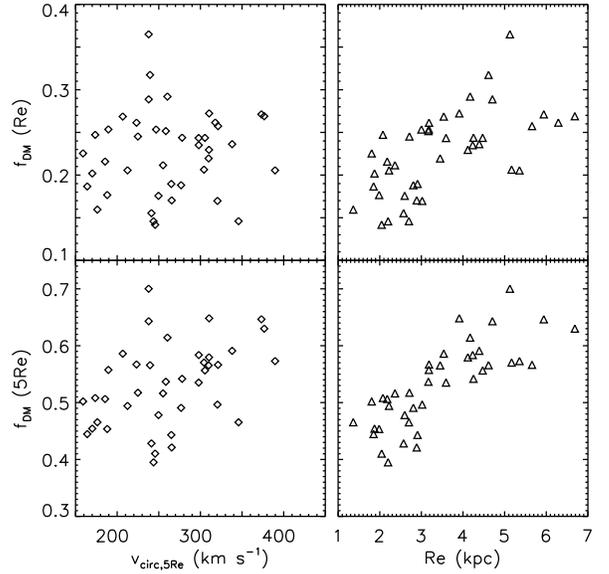}}
\caption{Dark matter fraction for all model galaxies at $1\re$ (upper panels) and $5\re$ (lower panels), versus outer circular velocity $v_{\rm circ}(5\re)$ (left panels) and versus effective radius (right panels).  Within $1\re$, the average mass in dark matter is about 25\% of the total, while within $5\re$, the average dark matter fraction is slightly above $50\%$. The dark matter fraction
  has little (at $5\re$) to no dependence (at $1\re$) on circular velocity, but increases clearly with $\re$.  }
  \label{dmfraction}
\end{center}
\end{figure}

Finally, we show the fraction of dark matter within 3-dimensional
spheres of $1\re$ (left panel of Fig.~\ref{dmfraction}) and $5\re$
(right panel) for the model galaxy sample. At $1\re$, the dark matter
fractions are between $15\%-30\%$, i.e., the luminous matter dominates
by a factor 3-7.  At $5\re$, the dark matter fractions are between
$40\%-65\%$, i.e., the amount of dark matter is now on average larger
than the mass in stars. We can infer from Fig.~\ref{dmfraction} that
the DM fraction increases with stellar mass (or circular velocity),
and that this increase is mostly due to a dependence on effective
radius rather than circular velocity \citep[see
also][]{Hilz_etal2012, Hilz_etal2013}. These  model values approximately agree with dark matter fractions inferred from power-law dynamical modelling of PNe and GC observations \citep[][Fig.~7]{Deason_etal2012} and strong lensing \citep[][Fig.~7]{Auger_etal2010}, for Salpeter IMF.

\subsection{Shape correlation and alignment}\label{shapecorr}

Observed ETGs have ellipticities up to $\epsilon\simeq 0.8$
\citep{Bernardi_etal2003, Krajnovic_etal2011}.  The distribution of
ellipticities depends on sample selection; it is approximately flat up
to ellipiticity $\epsilon\simeq 0.7$ for the recent ATLAS$^{3D}$ sample
\cite[see][Fig.~7]{Krajnovic_etal2011}.  The shapes of the outer dark
haloes ($\gtrsim 10\%~r_{\rm vir} $) can be estimated from the shear
patterns in weak gravitational lensing data \citep{Hoekstra_etal2004,Mandelbaum_etal2006, vanUitert_etal2012}.  Some of these studies
have found that the dark matter haloes of red galaxies on scales
beyond $\sim 0.1 r_{\rm vir}$ are preferentially aligned with the lens
galaxies, but the signal is not as clear as might have been expected.

Dark matter simulations predict that the haloes are triaxial
\citep{Jing_suto2002, Allgood_etal2006}. In hydrodynamical simulations
of disk galaxies, the inner haloes become preferentially aligned with
the disk and misaligned with the outer haloes \citep{Bailin_etal2005,
Bett_etal2010, Hahn_etal2010}, while in binary major mergers the
short axis of the stellar remnant is found to be oriented
perpendicular to the long axis of the surrounding halo
\citep{Novak_etal2006}.  It is therefore of interest to investigate
the shapes and orientations of the stellar and dark matter
distributions for the current model galaxies, which have a very
different formation history.

\begin{figure}{}
\begin{center}
\resizebox{9.cm}{!}{\includegraphics{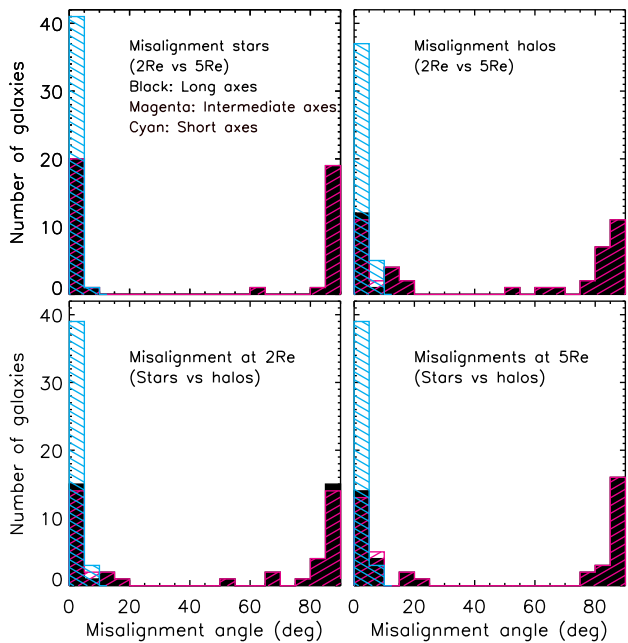}}
\caption{Alignment of the principal axes of the stellar and dark matter components. Histogram of misalignment angles for all 42 simulated galaxies, between: principal axes of the stellar component at $2\re$ and $5\re$ (top left); of the dark matter component at $2\re$ and $5\re$ (top right); of stars vs.\ dark matter at $2\re$ (bottom left); and stars vs.\ dark matter at $5\re$ (bottom right). Colours in each panel denote misalignment angle of the respective short axes (cyan), intermediate axes (magenta), and long axes (black).  }
\label{alignment}
\end{center}
\end{figure}

To study the alignment of the simulated galaxies and their host dark
matter haloes, we determine the principal axis directions of both
components at small radii ($2\re$) and large radii ($5\re$) at z=0. We
find these quantities by diagonalising the moment of inertia tensor
iteratively from all particles inside ellipsoids $x^2+(y/p)^2+(z/q)^2
= f^2\re^2$, where $f=2$ or $5$, $p=b/a$, $q=c/a$ and, e.g., the latter is
determined from $q\equiv c/a=\sqrt{{\tilde I_{zz}}/{\tilde I_{xx}}}$
where $\tilde I_{xx},~\tilde I_{zz}$ are the diagonalised moments of
inertia for all stellar particles within the ellipsoid. Because within
$1\re$ there are only a few hundred dark matter particles for many of
our models, we have chosen the inner radius at $2\re$ in order to
ensure reliable results.  The particle mass for the stellar component
is $\sim 1/5$ of the dark matter particle mass, so there is a
sufficient number of stellar particles even within quite small radii
to define the shapes of the `luminous' galaxies.

Figure~\ref{alignment} shows the histograms of misalignment angles
between the short, intermediate, and long axes of the stellar and halo
components at $2\re$ and $5\re$ for each component separately, and
between both components. The short axes (cyan) of the stars at
different radii, of the dark matter at different radii, and of the
stars and dark matter distributions are always very well aligned
(within $\lesssim 5^\circ$). The stellar minor axis
is therefore always perpendicular to the dark halo major axis.

The long axes (black) and intermediate axes (magenta) of the stars
and of the dark halo, or of either of these components at different
radii, are approximately aligned within ($\lesssim 15^\circ$) in
about half of the cases; in the other half of the cases they are
misaligned by roughly 90 degrees, meaning that the intermediate and
long axes have switched between different radii or components. This
can happen most easily when the system is nearly axisymmetric.

The cosmological sample contains simulated galaxies
formed in different ways, including some which underwent a gas-rich
major merger, but the evolution usually includes minor mergers as
well \citep{Oser_etal2012, Johansson_etal2012, Naab_etal2013}.
\citet{Novak_etal2006} show that in hydrodynamic simulations of
binary major mergers, the minor axis of the oblate stellar merger
remnant is generally perpendicular to the major axis of the
surrounding prolate-triaxial dark matter halo, due to the influence
of angular momentum and dissipation. Because the baryons cause the
inner halo to evolve towards oblate shape
\citep[e.g.][]{Kazantzidis_etal2004, Bailin_etal2005}, this would
partially explain our results.  In minor mergers, the stirring of
the pre-existing system by the incoming dark matter is similar for
the dark matter and stars. This may explain why dark halo and
stellar component are aligned at similar radii also for minor merger
dominated formation histories.

\begin{figure}{}
\begin{center}
\resizebox{9.cm}{!}{\includegraphics{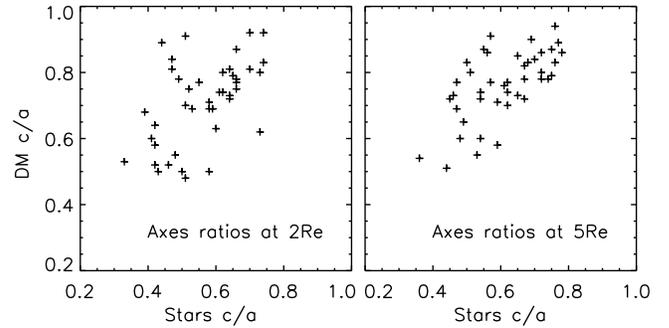}}
\caption{Short-to-long axis-ratios of stellar components versus halo
components at $2\re$ (left) and $5\re$ (right). The $c/a$ of stars
and haloes strongly correlate at large radii but less strongly
at small radii.}
\label{shape}
\end{center}
\end{figure}

\begin{figure*}{}
\begin{center}
\resizebox{9.2cm}{!}{\includegraphics{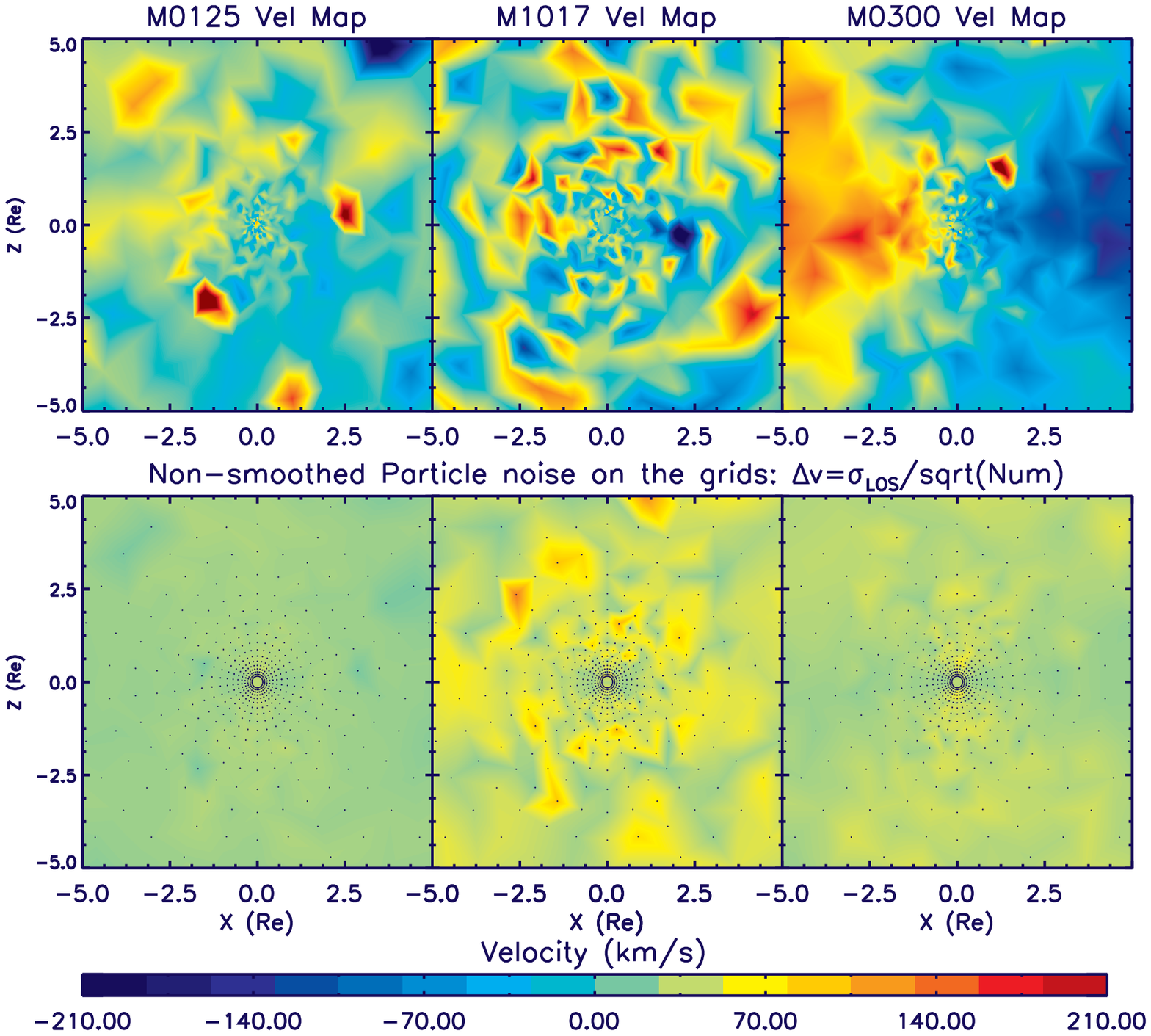}}\resizebox{9.2cm}
{!}{\includegraphics{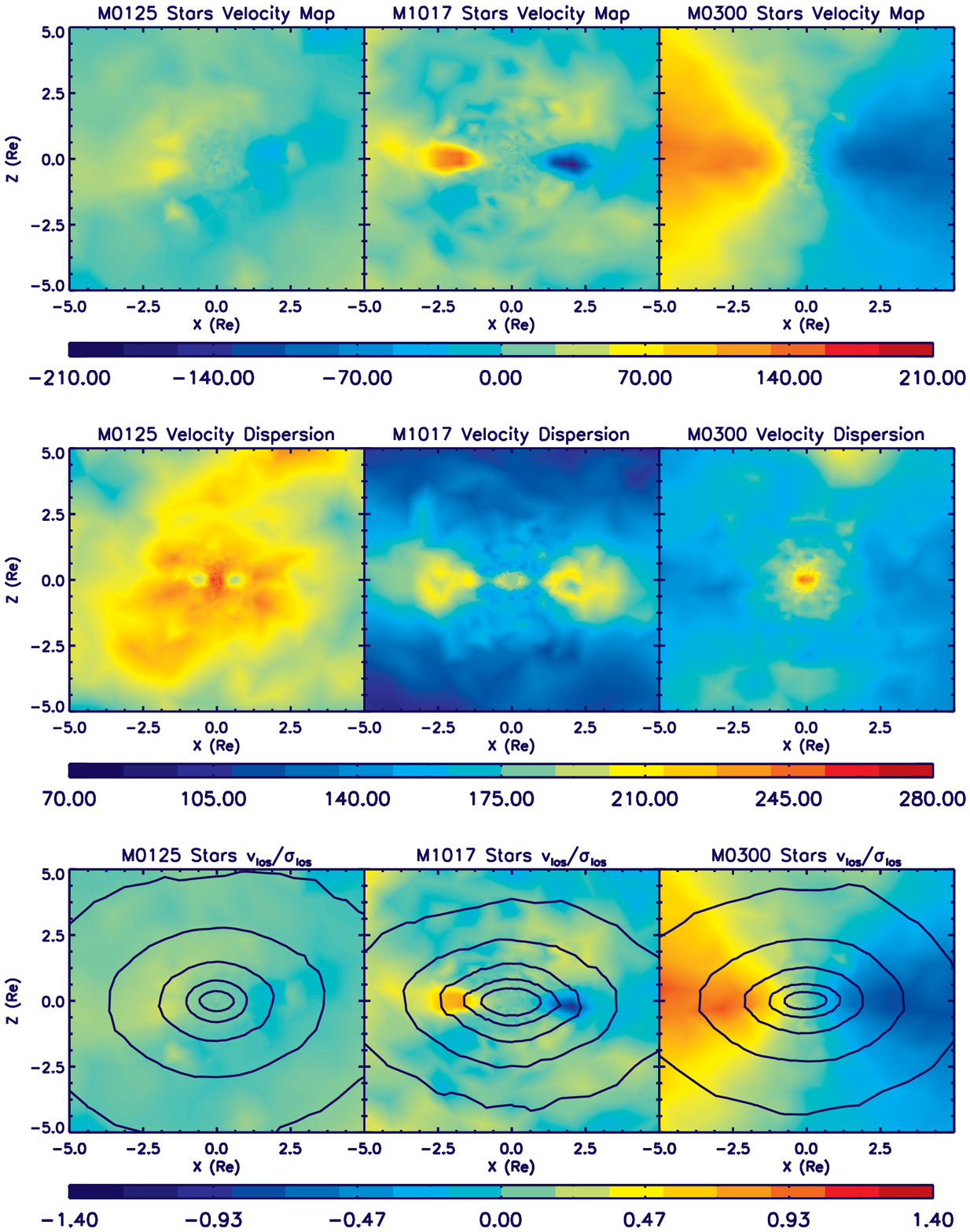}}
\caption{Edge-on \LOS\ kinematics of selected model galaxies. {Upper panels,
left column:} mean \LOS\ velocity maps for the stellar components of
three cosmological galaxies (labeled M0125, M1017, M0300),
computed from the particle distributions for the snapshot at $z=0$
\citep{Oser_etal2010}. The maps are interpolated from a grid in
radius and angle of $n_r\times n_\phi=20\times 30$ cells on a
$(10\re)^2$ region. Signatures from both particle noise and from a
number of small substructures are visible. {Lower panels, left
column:} particle noise error on the mean velocity for the same
galaxies. The grid points (black dots) are overplotted.  {Upper
panels, right column:} Temporally smoothed mean \LOS\ velocity maps
(${\bar v}$) for the stellar components of the same model galaxies.
{Middle panels, right column}: Maps of temporally smoothed
velocity dispersion $\sigma$.  {Bottom panels, right column:}
Maps of ratio ${\bar v}/\sigma$. The surface densities of the
stellar components are overplotted on these maps; contour levels are
$10^{7.0},10^{7.5},~10^{8.0},~10^{8.5},~10^{9.0} \Msun
\kpc^{-2} {\rm h}$.  The three model galaxies chosen here are: a slow rotator
(M0125), a slow rotator with a peak of $\lambda(R)$ around
$2\re$ (M1017), and a fast rotator with increasing $\lambda(R)$
up to $2\re$ (M0300); see Sect.~\ref{rotator}.  }\label{nosmooth}
\end{center}
\end{figure*}

Figure~\ref{shape} shows the short-to-long axis ratios of the dark
matter haloes versus those of the stars, at both $2\re$ and $5\re$.
These axis ratios where obtained by viewing the system along the
intermediate axis of the stellar component, and computing the
projected axis ratio $c/a$ from the diagonalised 2D projected moment
of inertia tensor.  At both $2\re$ and $5\re$, the axis-ratios of the
galaxies are in the range [0.4,0.8] while the haloes are slightly
rounder, $c/a \in [0.5,0.95]$.  From the figure we see a moderately
strong correlation between the shapes of the stellar and halo
components, with scatter in the halo axis ratio at given stellar $c/a$
of $\sim \pm 0.1$ at $5\re$ and $\sim\pm0.2$ at $2\re$.

In summary, the short axes of the simulated galaxies and their host
dark matter haloes are well aligned within $\lesssim 5^\circ$
througout the radial range probed ($2\re$-$5\re$), and their shapes
are correlated. Long and intermediate axes are either aligned or
misaligned by $90^\circ$, i.e., switch their order, within slightly
larger ($\lesssim 15^\circ$) scatter.

\section{Line-of-sight kinematics of simulated galaxies to large radii}\label{kin}
As is well-known, the outer kinematics of \ETGs\ are difficult to
measure because the stellar densities and surface brightness profiles
decrease rapidly at large radii.  The same problem also exists in the
simulated galaxies where the particles follow similar density
profiles.  In addition to Poisson noise, there is also a further
source of fluctuations that arises from various small satellites
around the central galaxies.

In order to investigate the effects of the fluctuations from both
low particle numbers and satellites, we consider three cosmological
galaxies from our sample of \citet{Oser_etal2010} more closely (see
also Figure \ref{densityfit}).
Velocity maps for these galaxies at $z=0$
are shown in Fig.~\ref{nosmooth}. The projection direction is along
the intermediate axis of the stellar distribution within one
effective radius\footnote{The
maps are interpolated from a polar grid as described in
Sect.~\ref{ics}.}. The unsmoothed snapshot maps of mean \LOS\ velocity
${\bar v}$ (upper panels in the left column of Fig.~\ref{nosmooth})
show significant fluctuations.  A map of the error $\delta {\bar v}$
in the mean \LOS\ velocity is shown in the lower left panels of
Fig.~\ref{nosmooth}. This is defined as $\delta {\bar v}_j=\sigma_j/\sqrt{N_j}$, where $\sigma_j$ and $N_j$ are the
velocity dispersion and number of particles in the $j_{\rm th}$ cell
grid for this line-of-sight. Typical values for these fluctuations are
$\delta {\bar v}_j=20-80~\kms$.  Model M1017 is an extreme case where
the error can reach almost $50\%$ of the mean \LOS\ velocity itself.

\begin{figure}{}
\begin{center}
\resizebox{9.2cm}{!}{\includegraphics{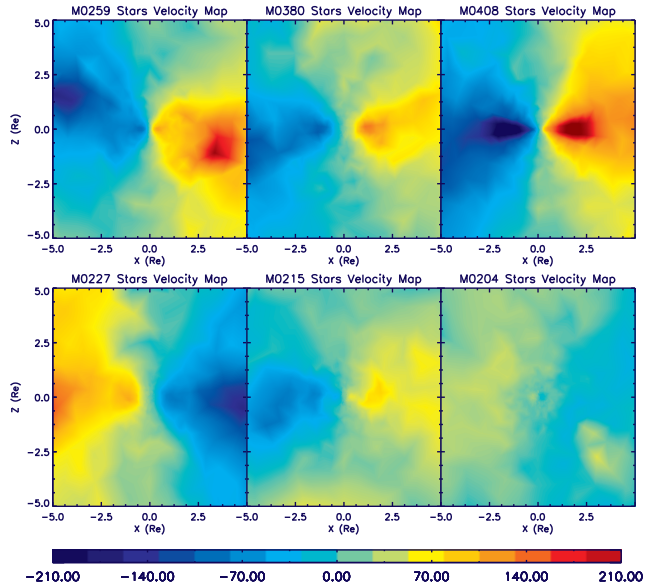}}
\caption{Edge-on mean \LOS\ velocity maps for the stellar components
of further 6 selected model galaxies. Upper panels show three
simulated galaxies with large scale rotation. The galaxy in the top
right panel has a pronounced disk component from a gas rich major
merger.  The lower panels show three simulated galaxies with a
dissipationless merging history. The galaxy in the bottom right
panel shows weak major and minor axis rotation; it has a formation
history with many minor mergers. }
\label{6gals}
\end{center}
\end{figure}

In addition to the fluctuations caused by particle noise, these
velocity maps show a number of well-defined substructures. These
satellites will be moving along their own orbits until they finally
merge with the central galaxy, and can locally have quite different
\LOS\ velocities from the host galaxy particles. Thereby the satellites
change the 1-dimensional \LOS\ velocity profiles and thus ${\bar v}$,
and they may also affect the angular momentum profiles. We have
decided not to take all these small satellites out, but to consider
the host galaxy after these substructures have phase-mixed away in its
large-scale gravitational potential. The NMAGIC code used for this
purpose (see Sect.~\ref{ics}) also computes a time average of the
observables ${\bar v}$ and $\sigma_j$ which allows us to obtain more
reliable \LOS\ kinematics at large radii than would otherwise be
possible.

Recomputing in this way the projected \LOS\ kinematics for the three
galaxies in Fig.~\ref{nosmooth} results in the kinematic maps
presented in the right column of Fig.~\ref{nosmooth}. The three maps
in the upper right panel are mean velocity maps ${\bar v}$.
Fluctuations are greatly reduced compared to the unsmoothed maps in
the left column. Also the fluctuations arising from the small
satellites are mostly smoothed out. The middle panel in the right
column of Fig.~\ref{nosmooth} shows maps of \LOS\ velocity dispersion
$\sigma$, and the lower panel shows maps of the ratio ${\bar
 v}/\sigma$ with overplotted stellar surface density contours.

For model M0125 (always in the left panels), the velocity field shows
a mild rotation, and its velocity dispersion map shows large $\sigma$
within $2.5 \re$, implying that this galaxy is pressure-supported. For
model M0300 (right panels), rapid disk-like rotation is seen from the
mean velocity map, while $\sigma$ is only large in the centre ($R<
0.5\re$); this object is significantly supported by rotation.
Finally, in model M1017 (middle panels), one can see a hot disk-like
structure at intermediate radii ($\sim 2.5\re$) in both the velocity
and $\sigma$ maps, while the rest of the galaxy has lower
dispersion. Model M1017 is mostly pressure-supported with a
rotationally supported component. While this model is somewhat
unusual, the other two are quite typical for the large radius
kinematics of fast and slow rotator galaxies in the simulated \ETG\
sample. Observed outer velocity fields similar to the three cases
shown here are those of NGC 5846 (a slow rotator galaxy), NGC 4564 (a
fast rotator with a disk-like velocity field at intermediate radii;
for both see \citealt{Coccato_etal2009}), and NGC 1316 (a rapidly
rotating merger remnant; see \citealt{McNeil-Moylan_etal2012}).

Figure~\ref{6gals} shows line-of-sight velocity fields for 6
additional simulated galaxies out to $5\re$. These galaxies have
been chosen as representatives for different dissipational and
dissipationless formation histories.  For example, the galaxy in the
top right panel has a pronounced disk component around $R\sim2\re$
from a gas rich major merger, and the galaxy in the bottom right
panel shows weak major and minor axis rotation; it has a formation
history with many minor mergers. For further
details about the formation histories of these simulated galaxies
see Naab et al.\ (2013, in preparation).

\subsection{Root Mean Square Velocity Profiles}\label{vrms-pro}
\begin{figure}{}
\begin{center}
\resizebox{6.cm}{!}{\includegraphics{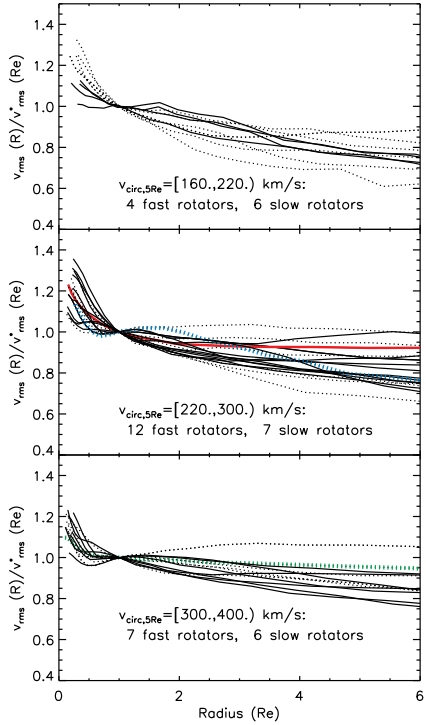}}
\caption{Radial profiles of root mean square
    line-of-sight velocity $v_{\rm rms}$ for the stellar components of
    all 42 simulated galaxies, for edge-on view and in three bins of
    $v_{\rm circ,5\re}$. The $v_{\rm rms}(R)$ profiles are
  normalised by their respective $v_{\rm rms}$ at $1\re$.  Solid lines
  are for fast rotators while dotted lines are for slow rotators; see
  Section~\ref{rotator}. Coloured curves depict the three galaxies
  with velocity fields shown in Fig.~\ref{nosmooth}: M0125 (green
  dotted line, bottom panel), M1017 (blue dotted line, middle panel),
  and M0300 (red solid line, middle panel).}\label{vrms_scaled}
\end{center}
\end{figure}

Observations of outer \ETG\ kinematics using planetary nebula velocities
have shown that most \ETGs\ are characterised by slowly declining
profiles of circularly averaged \RMS\ \LOS\ velocity $v_{\rm
  rms}(R)$, but with a significant minority of galaxies for which the
\RMS\ velocity declines rapidly \citep{Coccato_etal2009}. It is
therefore of interest to investigate the equivalent radial profiles
for the (re)simulated galaxies considered here.

Figure~\ref{vrms_scaled} shows the circularly averaged profiles of
$v_{\rm rms}(R)$ for the 42 galaxies from \citet{Oser_etal2010} in the
three usual bins of $v_{\rm circ}(5\re)$. The viewing direction for
all simulated galaxies is edge-on along the intermediate axis of the
stellar distribution. The $v_{\rm rms}(R)$ include projected rotation
and velocity dispersion of all star particles, and are normalised by
the respective values of $v_{\rm rms}(1\re)$.  We find that the
$v_{\rm rms}(R)$ profiles decline moderately with radius for the
low-mass systems, whereas they decrease only mildly for the high-mass
systems. This agrees with the CVCs shown in Fig.~\ref{vc_vcbin}. The
$v_{\rm rms}(R)$ profiles of fast and slow rotators (see
Sect.~\ref{rotator}) are not significantly different.

The $v_{\rm rms}(R)$ profiles in Fig.~\ref{vrms_scaled} are consistent
with the major group in \citet[][Fig.~15]{Coccato_etal2009}, but there
is no equivalent in the simulations for the rapidly falling profiles
shown in that paper (e.g., for NGC 3379).  For this comparison, we
also studied the $v_{\rm rms}(R)$ profiles of the simulated galaxies
in face-on projection, with \LOS\ parallel to the shortest axis. The
face-on profiles are similar to those in edge-on projection; at least
for the current sample of cosmological galaxies, the inclination angle
does not seem to be an important parameter for $v_{\rm rms}(R)$.
However, to follow the formation of thin disks in
large-volume cosmological simulations is still problematic, because
of the required high resolution \citep[see,
e.g.,][]{Guedes_etal2011,Brook_etal2012}, and this may also impact
the properties of simulated merger remnants from such disks. We note that binary merger simulations with orbits
drawn from cosmological simulations can result in steeply falling
velocity dispersion profiles in $\Lambda$CDM halos
\citep{Dekel_etal2005}.  Therefore in the context of cosmological
simulations this question should be revisited once the resolution
and the physical modelling is sufficient for modelling well-resolved
populations of disk and elliptical galaxies in a large volume.

\begin{figure}{}
\begin{center}
\resizebox{8.cm}{!}{\includegraphics{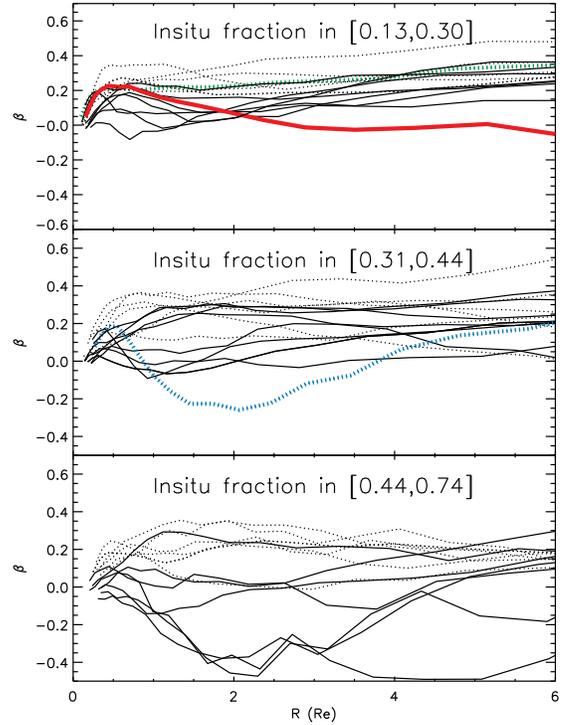}}
\caption{Velocity anisotropy profiles for the simulated galaxies in
bins of in situ fraction of stars within $10\% r_{\rm vir}$.
Line-styles and coloured curves are as described in
Fig.~\ref{vrms_scaled}. Galaxies with low in situ fractions have
more radially anisotropic orbit distributions, while tangential
anisotropy is seen only for systems with high in situ fraction.}
\label{beta}
\end{center}
\end{figure}

\subsection{Anisotropy profiles}

In these cosmological simulations, the stars in the model galaxies
have two origins: the inner in situ component forms at early times in
a period of rapid star formation, while the second component stems
from accreted stellar particles which is predominantly but not only
found at large radii \citep{Oser_etal2010}.  The accreted component is
characterised by radially anisotropic velocity dispersions
\citep{Abadi_etal2006, Hilz_etal2012} because the merging satellites
come in on predominantly radial orbits, and so many of the stars
stripped from the satellites enter the host galaxy on nearly radial
orbits also.

\begin{figure*}{}
\begin{center}
\resizebox{17.cm}{!}{\includegraphics{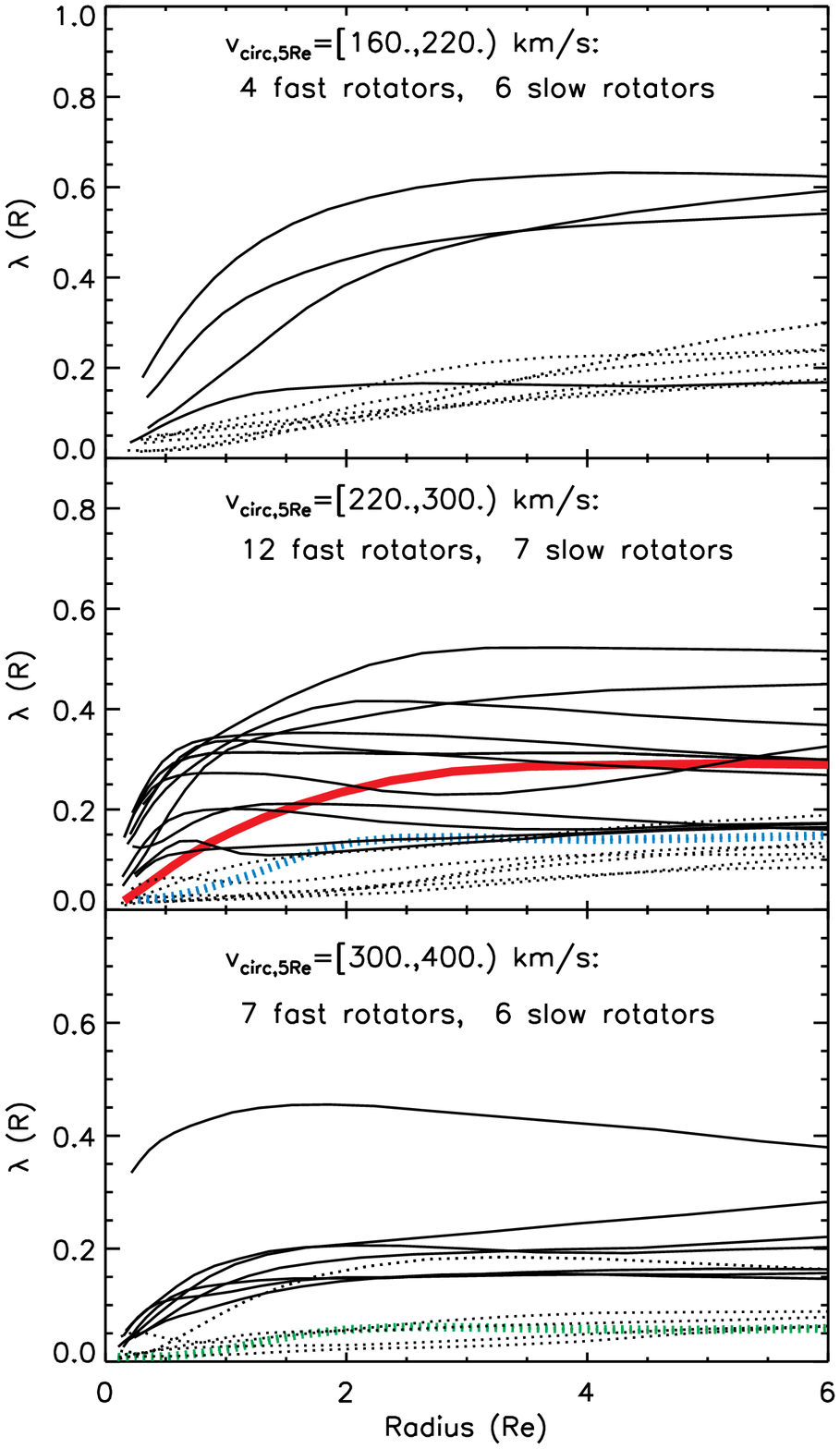}
                     \includegraphics{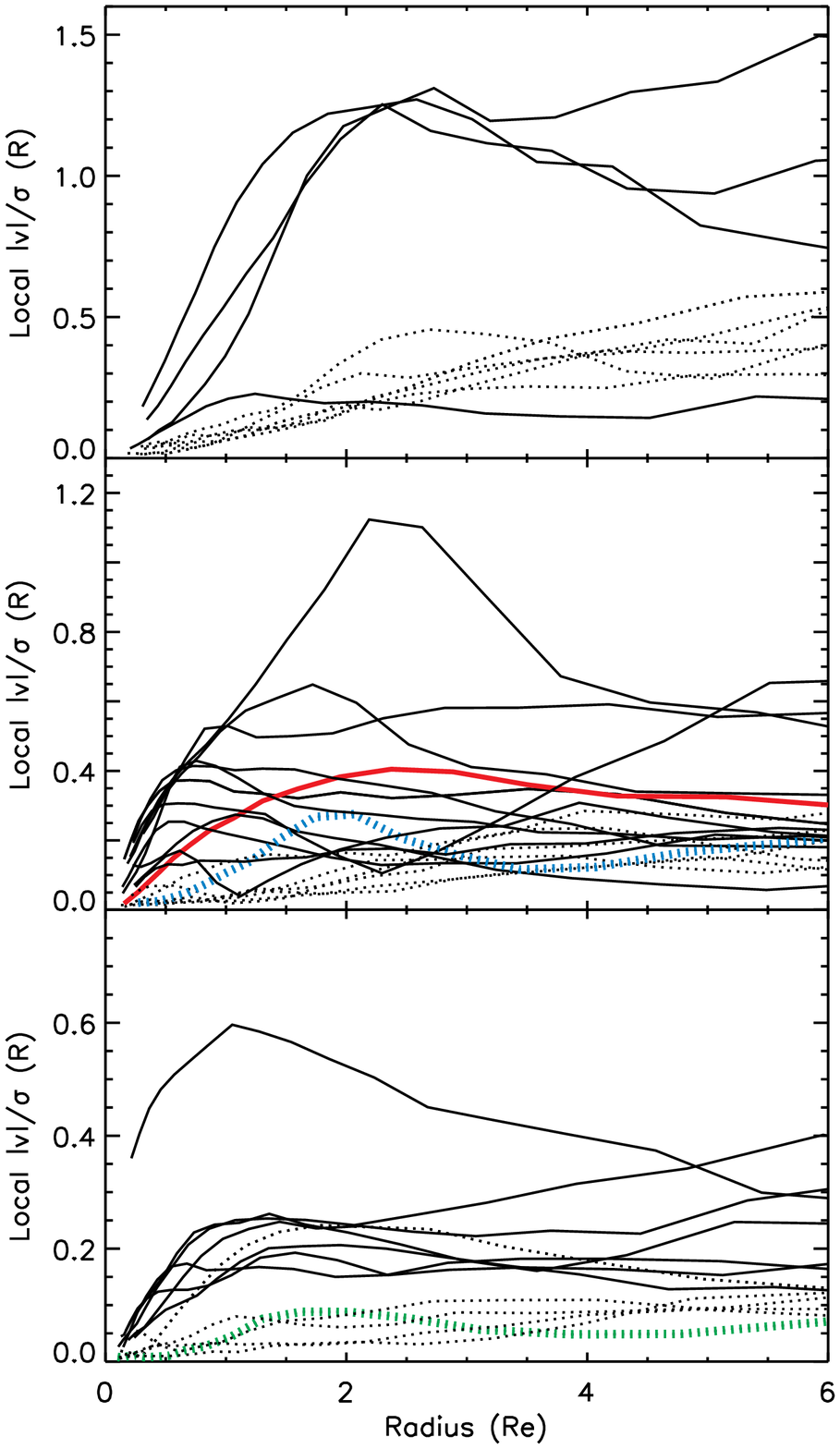}
                     \includegraphics{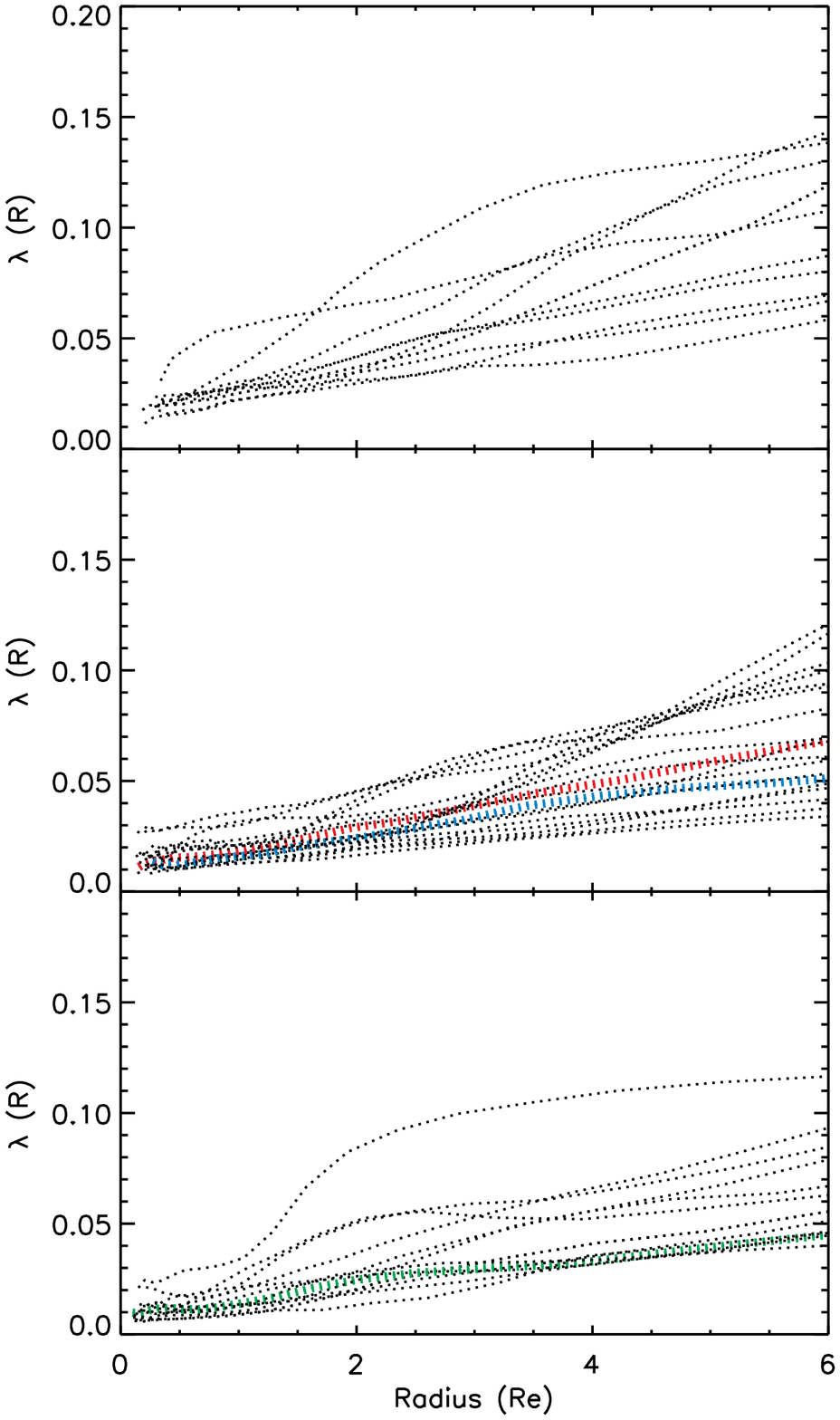}}
\caption{Specific angular momentum parameter $\lambda(R)$ profiles for
the sample of 42 cosmological re-simulation galaxies, binned by
circular velocity $v_{\rm circ}(5\re)$ and plotted outside the
softening radius $0.4~\kpch$.  Edge-on profiles (inclination
$i=90^\circ$) are shown in the left panels and face-on profiles
($i=0^\circ$) in the right panels.  The middle panels show $|{\bar
  v}|/\sigma$ profiles for $i=90^\circ$ in the same bins; these
provide more local rather than cumulative information on the
rotational properties. The solid lines are for fast rotators whereas
the dotted lines show slow rotators; this classification depends on
viewing angle.  The coloured curves point to three typical galaxies:
M0125 (green dotted lines on bottom panels), M1017 (blue dotted
lines on middle panels) and M0300 (red solid lines on middle
panels).  }\label{lambda_vcbin}
\end{center}
\end{figure*}

Therefore, we would expect the kinematics of model galaxies to be more
or less radially anisotropic depending on the relative fractions of in
situ and accreted stars. Figure~\ref{beta} shows anisotropy profiles
$\beta(r) \equiv 1 - (\sigma_\theta^2+\sigma_\phi^2)/2\sigma_r^2$ for
all model galaxies, divided into bins of in situ fraction. As
expected, tangentially anisotropic model galaxies can only be found in
the group with large fraction of in situ stars (lower panel and also
M1017 in the middle panel). Almost all simulated galaxies with low in
situ star fraction (upper panel) are radially anisotropic with
$\beta\simeq 0$-0.3; but many of the systems with higher in situ
fraction have similar anisotropies. As a result, there is no
correlation between both quantities for the whole sample.  We also
find that slow rotators are radially anisotropic (except for M1017),
while fast rotator model galaxies have both radially and tangentially
biassed anisotropy profiles.

For the radially anisotropic galaxies in
Figure~\ref{beta}, the $\beta(r)$ profiles are almost independent of
radius with typical values of $\beta\simeq0.1-0.3$ for $R>2R_e$.
Together with the CVCs in Fig.~\ref{vc_vcbin} this results in gently
falling $v_{\rm rms}(R)$ profiles as in Fig.~\ref{vrms_scaled},
according to the Jeans equation. To reproduce the steeply falling
velocity dispersion profiles of the quasi-Keplerian galaxies like
NGC 3379 often requires rising anisotropy profiles
\citep{Morganti_etal2013}, such as may arise from binary mergers of
gas-rich disk galaxies \citep[e.g.][]{Dekel_etal2005}.

\section{Angular Momentum profiles}\label{rotator}

\subsection{$\lambda(R)$ profiles}\label{sec-lambda}

To characterize the specific angular momentum of \ETGs,
\citet{Emsellem_etal2007} introduced the $\lambda(R)$ profile
$\lambda(R)=<R v>/<R\sqrt{v^2+\sigma^2}>$, which is a
luminosity-weighted, cumulative measure of projected angular momentum
per unit mass within radius $R$. They found that \ETGs\ can be divided
into fast and slow rotators, according to whether $\lambda(R=R_e)>0.1$
resp.\ $<0.1$. Analyzing simulated merger remnants,
\citet{Jesseit_etal2009} found that $\lambda(R)$ is a good proxy for
the true angular momentum of these remnants.
\citet{Emsellem_etal2011} revised the criterion for separating slow
and fast rotators to include ellipticity values at $R_e$ or $R_e/2$,
and they also considered the effect of inclination and an assumed
anisotropy on the resulting $\lambda(R)$ profile.

Much of the angular momentum of \ETGs\ could reside at large radii,
where kinematic measurements are more difficult than in the bright
centers. \citet{Coccato_etal2009} studied the kinematics of the
outer haloes of a sample of \ETGs\ with PNe.  They found (their
Fig.~14) that the $\lambda(R)$ profiles to large radii for the most
part confirm the separation into fast and slow rotators based on the
inner kinematics, but with some slow rotators having up to
$\lambda(R)\sim 0.3$ at large $R$ and some fast rotators whose
$\lambda(R)$ profiles decrease outwards.

It is therefore of some interest to compare the specific angular
momenta of the simulated galaxies with those of observed \ETGs, out
to several effective radii.  To obtain an overview of the angular
momentum properties of the simulated galaxies, we study the
$\lambda(R)$ profiles for all the 42 cosmological galaxies, and to
reach large radii we use the temporally smoothed observables
(Sect.~\ref{ics}) in the computation of $\lambda(R)$,
\beq\label{lambda}
\lambda(R_i)={\sum_{k=1}^{i} \sum_{j=1}^{N_{\phi}} m^P_{j,k} R_{k}|v_{j,k}|
             \over \sum_{k=1}^{i} \sum_{j=1}^{N_{\phi}}
                           m^P_{j,k} R_{k} \sqrt{v_{j,k}^2+\sigma_{j,k}^2}},
\eeq where the summation is over the kinematic grid. Eq.~9 is the analogue of the flux-weighted sums over
Voronoi pixels in \citet[][Eq.~6]{Emsellem_etal2007}, and of the
number-weighted sums in \citet[][Eq. 13]{Coccato_etal2009}.

Figure~\ref{lambda_vcbin} shows the resulting $\lambda(R)$ profiles,
again with the simulated galaxies binned according to their outer
circular velocity $v_{\rm circ, 5\re}$.  The left panels of
Fig.~\ref{lambda_vcbin} show the $\lambda(R)$ profiles for edge-on
view, while the right panels show the face-on $\lambda(R)$ profiles.
Most of the simulated profiles have a very regular form, reaching a
nearly constant value at $\sim 1-2R_e$, but some increase or decrease
more noticeably at large radii.

The $\lambda(R)$ profiles for a large number of nearby
ETGs are shown out to $\lesssim 1.5\re$ in Fig.~5 of
\citet{Emsellem_etal2011}. More extended $\lambda(R)$ profiles for
ETGs are shown in \citet[][out to $3~\re$]{Proctor_etal2009} and
\citet[][out to $9~\re$]{Coccato_etal2009}.  The variations and
general shapes of the $\lambda(R)$ profiles from the cosmological
galaxies agree with the observed $\lambda(R)$ profiles of \ETGs\ to
large radii, shown in Fig.~14 of \citet{Coccato_etal2009},
and the typical values at $5R_e$, $\sim 0.1$-$0.7$, also
agree.  Their slopes at small radii appear somewhat shallower
than the typical slopes seen in the SAURON and ATLAS$^{3D}$ data
\citep[e.g., Fig.~5 in][]{Emsellem_etal2011}. However, even
for the unusual model M1017, within $1.5\re$, there are similar
$\lambda(R)$ profiles in \citet{Emsellem_etal2011}.

As mentioned above, \ETGs\ can be classified into fast and slow
rotators according to their $\lambda(R)$ values at $\re$. For
classifying the simulated galaxies we follow
\citet{Emsellem_etal2007}, taking $\lambda(\re)>0.1$ to define fast
rotators (solid lines in Fig.~\ref{lambda_vcbin}) and
$\lambda(\re)<0.1$ for slow rotators (dotted lines). The main reason
for this choice is that this classification does not depend on
ellipticity. We find that among 42 cosmological galaxies, 23 are fast
rotators (based on the edge-on profiles) and 19 are slow rotators. The
fraction of slow rotators is largest in the group of simulated
galaxies with the highest circular velocities at $5R_e$.  In the
face-on view (right panels of Fig.~\ref{lambda_vcbin}), none of these
systems rotates rapidly, and they are all classified as slow rotators.
The fraction of slow rotators among the simulated
galaxies ($45\%$) is significantly more than in the ATLAS$^{3D}$
sample \citep{Emsellem_etal2011}; however, the sample selection
(predominantly massive galaxies in the simulation) is not
comparable to the ATLAS$^{3D}$ sample \citep[see][]{Oser_etal2010}.

It is clear from Fig.~\ref{lambda_vcbin} that most of the simulated
edge-on $\lambda(R)$ profiles are flat to slightly rising for
$R\gtrsim2\re$. Thus, most fast rotators at $R_e$ continue to rotate
rapidly at large radii. Most of the slow rotators have mildly
increasing $\lambda(R)$ in the outer regions, especially for the group
of less massive systems with circular velocities at large radii
$<220\kms$.

\subsection{$|{\bar v}|/\sigma (R)$ profiles}
To obtain a more detailed picture of the angular momentum in the
outer stellar haloes, it is useful to study the more local angular
momentum parameter $|{\bar v}|/\sigma(R)$ as well. The local $|{\bar
  v}|/\sigma$ profiles for all simulated galaxies in edge-on view are
shown in the middle panels of Fig.~\ref{lambda_vcbin}.  For most
simulated galaxies, the $|{\bar v}|/\sigma(R)$ profiles are nearly
flat or increasing mildly with radius after reaching a plateau at
$1-2R_e$. However, there are also a few exceptions for which $|{\bar
  v}|/\sigma(R)$ decreases at large $R$, and some with a strong local
concentration of specific angular momentum. These features do not
show up as well in the cumulative $\lambda(R)$ profiles.

\begin{figure}{}
\begin{center}
\resizebox{9.0cm}{!}{\includegraphics{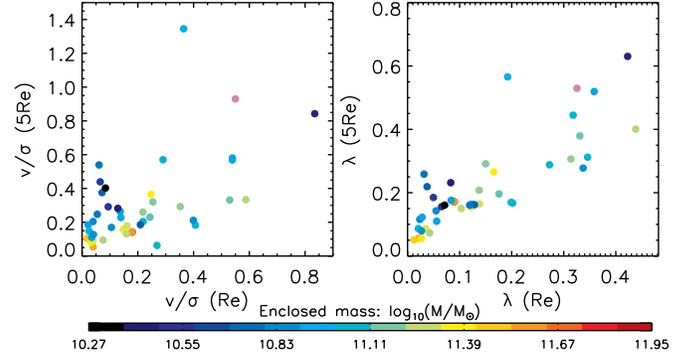}}
\caption{Relation between specific angular momentum parameters of the
outer stellar haloes and central regions for the 42 simulated
galaxies.  Left panel: local $|{\bar v}|/\sigma~(5\re)$ versus
$|{\bar v}|/\sigma~(\re)$, for edge-on view. Right panel:
$\lambda(5\re)$ versus $\lambda(\re)$. Stellar halo rotation
correlates with central rotation, especially for the flattest
galaxies. The colour code in both panels shows stellar mass enclosed
within $10\%~r_{\rm vir}$. }\label{angmoms_re5re}
\end{center}
\end{figure}

The left panel of Fig.~\ref{angmoms_re5re} shows a correlation with
much scatter between the central and outer local specific angular
momentum, $|{\bar v}|/\sigma$ at $1\re$ and $5\re$. The right panel of
Fig.~\ref{angmoms_re5re} shows the close relation between the
$\lambda(R)$ values at $1\re$ and $5\re$, which follows from the
nearly flat $\lambda(R)$ profiles. The correlation between the two
$\lambda(R)$ values is stronger than for the local angular momentum
$|{\bar v}|/\sigma$ at the same radii, because of the
cumulative nature of $\lambda(R)$ within radius $R$.
\subsection{Temporal smoothing needed for angular
momentum profiles at large radii}

In Fig.~\ref{nosmooth} we showed the velocity fields of three typical
galaxies: a slow rotator with very little rotation at any radii
(M0125, left panels), a fast rotator for which rotation is significant
at all radii (M0300, right panels), and a system rotating slowly in
the inner $\re$ but rotating rapidly for $R\sim 2-4 \re$ (M1017,
middle panels). Especially the ${\bar v}/\sigma$ maps in
Fig.~\ref{nosmooth} show the different rotational properties of these
model galaxies well.

\begin{figure}{}
\begin{center}
\resizebox{6cm}{!}{\includegraphics{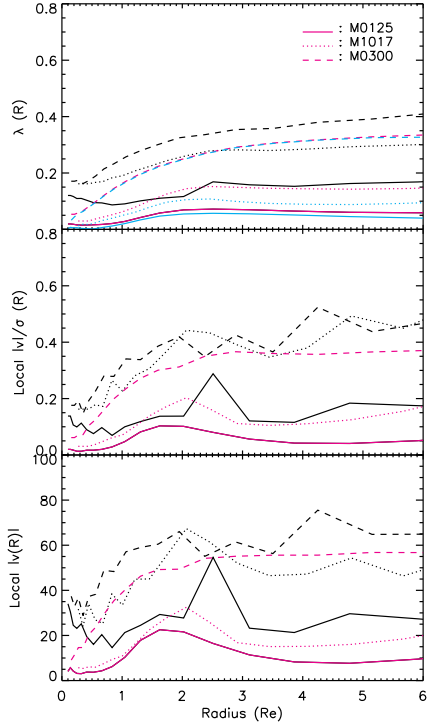}}
\caption{The 1-dimensional profiles of mean absolute velocity (lower
panel), local rotation parameter $|{\bar v}(R)|/\sigma$ (middle),
and cumulative specific angular momentum parameter $\lambda(R)$
(upper panel), for the three simulated galaxies M0125 (solid lines),
M0300 (dashed lines) and M1017 (dotted lines). Black curves show
profiles computed from unsmoothed data, magenta curves show
the results based on temporally smoothed data, and light-blue curves
show profiles based on two-sided averaging (see text).}\label{lambda_3gals}
\end{center}
\end{figure}

We use the same three simulated galaxies to show the beneficial effect
of the temporal smoothing on the cumulative $\lambda(R)$ and local
$|{\bar v}(R)|/\sigma$ profiles in Fig.~\ref{lambda_3gals}, where the
smoothed profiles from Fig.~\ref{lambda_vcbin} are shown in red and
the unsmoothed profiles are overplotted in black, for comparison.  The
upper panel shows that the amplitudes of the unsmoothed $\lambda(R)$
profiles are significantly higher than those of the time-smoothed
counterparts.  The lower panels show a similar effect in the local
$|{\bar v}(R)|$ and $|{\bar v}|/\sigma(R)$ profiles; these have large
fluctuations for a single snapshot, especially for the slow rotators,
with values of up to a factor of $2$ larger than in the smoothed
profiles.

The reason for the higher amplitudes in the unsmoothed case lies in
the definition of the $\lambda(R)$ parameters in terms of absolute
values of mean velocity, so that negative and positive fluctuations
cannot cancel in the angular or radial summation.  The unsmoothed
profiles can therefore be biased significantly by particle noise,
especially for the slow rotators (such as model M0125) and in the
center, and by global asymmetries (such as seen in model M0300 out to
large radii). The light-blue curves in the top panel of
Fig.~\ref{lambda_3gals} show the results of an additional test for
this effect. They are determined from computing $\lambda(R)$
separately for the positive and negative velocity sides of the mean
velocity maps in Fig.~\ref{nosmooth}, but without the absolute value
convention of eq.~\ref{lambda}, and then adding the absolute
$\lambda(R)$ values from both sides. This has the effect that positive
and negative velocities on each side are allowed to cancel, which
mimicks the effect of removing the noise or the asymmetry by the
phase-mixing that occurs during temporal smoothing.  It is clear from
Fig.~\ref{lambda_3gals} that the profiles obtained from this
asymmetric averaging and from temporal smoothing agree closely with
each other for all three simulated galaxies. This shows that to obtain
a correct indication of the angular momentum of the system, the time
averaging is necessary.

\subsection{Correlations of angular momentum and ellipticity}
\begin{figure}{}
\begin{center}
\resizebox{9.0cm}{!}{\includegraphics{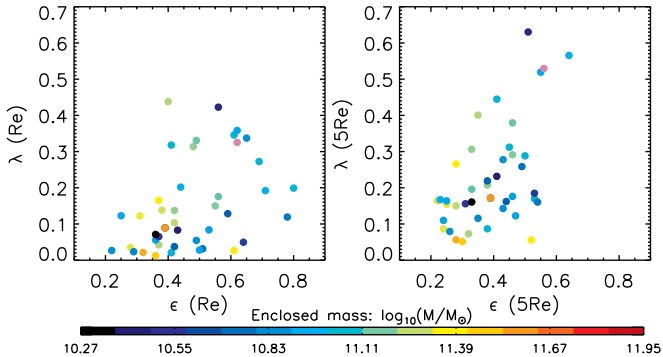}}
\caption{Relation between rotation, ellipticity and mass for outer
stellar haloes and central regions of the 42 simulated galaxies:
$\lambda(R)$ versus ellipticity $\epsilon$,
at $\re$ (left) and $5\re$ (right), for edge-on view. The colour
code in all panels shows stellar mass enclosed within $10\%~r_{\rm vir}$. The more massive galaxies are rounder and have lower
specific angular momenta, while the less massive galaxies are more
evenly distributed in both parameters.}\label{rotator_re5re}
\end{center}
\end{figure}

The observations of the SAURON \citep{Cappellari_etal2007} and
ATLAS$^{3D}$ projects \citep{Emsellem_etal2011} have shown evidence
for correlations between specific angular momentum parameters (either
$|{\bar v}|/\sigma$ or $\lambda(R)$ profiles) with ellipticity and
mass of these \ETGs.  S0-like fast rotator ETGs agree with being a
family of oblate systems viewed at random inclination angles, while
slow rotator ETGs are moderately flattened ($\epsilon\simeq 0.3$) and
often show kinematical misalignments, indicating that they are
triaxial systems.

To compare with these observations, we show in
Figure~\ref{rotator_re5re} the $\lambda(R)$ parameters versus the
ellipticities $\epsilon$ of the 42 simulated galaxies, both at $1\re$
(left) and $5\re$ (right).  Here ellipticity $\epsilon \equiv 1-c/a$,
where $a, c$ are the long and short semi-axes of the projected ellipse
in edge-on projection (see \S\ref{shapecorr}).  We list the values of
ellipticity at $1\re$ and $5\re$ in columns $3$ and $4$ of Table
\ref{fitting}, recalling that the central ellipticities for the lower
mass systems have some uncertainty due to softening.
The values of $\epsilon$ and $\lambda$ at $1\re$ agree with those
from direct analysis given in \citet{Naab_etal2013}, but small
differences exist which can be attributed to the different analysis
procedures. Some of the galaxies contain highly flattened
disk-like structures which result in large ellipticities at
$R\sim\re$.

Generally, the simulated slow rotators are rounder than the fast
rotators, similar as in the observations. The distributions at $1\re$
and $5\re$ are similar, despite the already mentioned possibility that
the ellipticity within $\re$ could be somewhat influenced by
softening.  The left panel of Fig.~\ref{rotator_re5re} is in general
agreement with the ATLAS$^{3D}$ observations \citep[Fig.~6
in][]{Emsellem_etal2011}; however, the ellipticity range is narrower
for the simulated galaxies and there are no
simulated galaxies with $\lambda(R_e)>0.5$. This is because the
ATLAS$^{3D}$ sample contains some very flat \ETGs\ which may be disk
galaxies, while our sample galaxies are mostly spheroidal
galaxies. The equivalent face-on distribution is
compressed towards $\lambda(R_e)=0$, $\epsilon(R_e)=0$ and values
$\lambda(R_e)>0.15$, $\epsilon(R_e)>0.4$ are rare. A
randomly projected sample fills the same area as in
Fig.~\ref{rotator_re5re} but has a stronger weighting towards
the origin of the plot \citep[see][Fig.~11]{Naab_etal2013}.

The colour bar in Fig.~\ref{rotator_re5re} labels the stellar mass of
the simulated galaxies within $10\%~r_{\rm vir}$. This shows the
correlations between angular momentum, ellipticity and stellar mass
for the simulated galaxies:  more massive galaxies
rotate more slowly on average (have smaller $\lambda(R)$) and
they are rounder, while the specific angular momenta for the less
massive galaxies are spread over a wider range, in broad agreement
with Fig.~11 in \citet[][]{Emsellem_etal2011} and Fig.~16 in
\citet[][]{Coccato_etal2009}.

\section{Conclusions and discussion}\label{conc}

In this work, we presented the first detailed
analysis of the inner and outer dynamics of a large sample of
simulated galaxies, studying the mass distributions, outer
kinematics, and angular momentum distributions of 42 resimulated
galaxies from the high-resolution cosmological re-simulations of
\citet{Oser_etal2010}. Here we summarize and discuss our main
results.

1. The stellar components of the model galaxies have approximately
power-law density profiles well fitted by cored S\'ersic profiles with
large $n$, falling somewhat less steeply towards large radii than \ETGs\
\citep[e.g.,][]{Kormendy_etal2009}.

2. Due to the interaction between dark matter and baryons during the
assembly processes, the dark matter density profiles of the model
galaxies deviate from the NFW-like profiles characteristic for dark
matter only models in the inner few 10's of kpc \citep{NFW1996,
  Navarro_etal2010}.  The DM density slopes and the slopes of the DM
CVCs for $r\in[(1-2)\re, 10\re]$ agree with power laws and vary
systematically with mass, such that the DM CVC is approximately flat
($R^{0}$) for less massive systems and slightly rising ($R^{0.3}$) for
high mass galaxies.

3. The corresponding total CVCs are slightly falling ($R^{-0.3}$)
for the less massive systems and approximately flat ($R^{0}$) for
the more massive model galaxies. This agrees with mass
determinations for local galaxies (see Introduction).

4. The dark matter fractions within the projected stellar half
  mass radius $\re$ are in the range $15$-$30\%$ and increase to
  $40$-$65\%$ at $5\re$. Larger and more massive galaxies have higher
  dark matter fractions. The fractions and trends with mass and size
  are in agreement with observational estimates.

5. The short axes of the simulated galaxies and their host
dark matter haloes are well aligned within $\lesssim 5^\circ$
througout the radial range probed ($2\re$-$5\re$), and their shapes
are correlated. Long and intermediate axes are either aligned or
misaligned by $90^\circ$, i.e., switch their order, within slightly
larger ($\lesssim 15^\circ$) scatter.

6. We computed mean velocity ${\bar v}$, velocity dispersion $\sigma$,
and local ${\bar v}/\sigma$ fields out to $5\re$ for the simulated
galaxies and illustrated their kinematic diversity.  We temporally
smoothed these velocity fields in order to suppress particle noise and
fluctuations from small satellites and illustrated the necessity and
effect of this in some detail.  The simulated galaxy sample contains
both purely dispersion-supported systems with no or little rotation,
and objects that show disk-like rotation at a level of $v/\sigma\simeq
1$. We also showed a rarer case with a disk-like structure at $\gtrsim
2.5\re$.  The observed outer velocity fields of galaxies from PNe show
a similar kinematic diversity.

7. Radial profiles of root mean square velocity $v_{\rm rms}(R)$ are
slowly declining, independent of whether the simulated galaxies are
fast or slow rotators, and similar for edge-on and face-on
projections.  These profiles resemble the majority group of outer
$v_{\rm rms}(R)$ profiles determined in nearby \ETGs\ from PNe
velocities by \citet{Coccato_etal2009}; however, there are no
analogues in the simulated galaxies for the rapidly falling $v_{\rm
  rms} (R)$ profiles seen in the observed sample. This could mean that
these objects form through a channel different from the simulated
galaxies studied here, or it could be related to the difficulty of
modelling disk galaxies in the cosmological simulations; this issue
requires further study.

8. We determined cumulative $\lambda(R)$ and local $v/\sigma (R)$
angular momentum parameter profiles for the stellar components of
all simulated galaxies from the time averaged velocity fields.  For
most simulated galaxies, the edge-on $\lambda(R)$-profiles are flat
or slighty rising within $2\re-6\re$. Most fast rotators rotate fast
at both small and large radii, but some have decreasing rotation at
large radii. Lower mass slow rotators have mildly increasing
$\lambda(R)$ with $\lambda(6\re) \in (0.1,0.3]$, whereas high-mass
slow rotators have flat $\lambda(R)$-profiles.  Overall, $\lambda$
increases with ellipticity, but with much scatter.  These properties
appear to broadly agree with those of observed \ETGs.

9. Simulated galaxies with a large fraction of accreted stars are
generally radially anisotropic. Only systems with a high fraction of
in-situ stars show tangential anisotropy. These trends are due to the
fact that in the simulation the accreted stars are tidally dissolved
from merging satellites on preferentially radial orbits.  We also find
that massive galaxies and slow rotators amongst the simulated
galaxies are mostly radially anisotropic, while the fast rotators have
both radial and tangential anisotropy profiles.

\section{Acknowledgments}

This project was mostly carried out while XW was as a postdoc in the
dynamics group at MPE. During the final stages she acknowledges an
Alexander von Humboldt fellowship at AIfA at Bonn University.  We
thank F.~de Lorenzi, P.~Das, and L.~Morganti for their work and
help with the NMAGIC code, and the anonymous referee for comments
which helped to improve the paper.

\appendix

\begin{table}
\begin{center}\vskip 0.00cm
  \caption{Parameters for stellar projected half mass radii,
    ellipticities, and rotation parameters for the simulated galaxy
    sample: designations of the model galaxies ($1${\it st} column),
    the $2${\it nd} column is $\re$ in $\kpch$, i.e., the effective
    radius containing half the projected stellar mass within
    $10\%~r_{\rm vir}$; 3{\it rd}-4{\it th} columns give ellipticities
    at $1\re$ and $5\re$ for the stellar mass distributions, 5{\it th}
    and 6{\it th} columns give rotation parameters $\lambda(5\re)$,
    and ratio $R_v\equiv [v/\sigma(5\re)]$/[$v/\sigma(\re)$]. }
\begin{tabular}{lrcccccccc}
\hline
Model &  $R_e$& $\epsilon~(\re)$ & $\epsilon~(5\re)$
& $\lambda(5\re)$ & $R_v$\\
\hline
M0094  &        3.86   &    0.39 &    0.39 &    0.17 &    0.80 \\
M0125  &        4.81   &    0.32 &    0.28 &    0.06 &    1.29 \\
M0175  &        4.28   &    0.36 &    0.30 &    0.05 &    6.61 \\
M0190  &        4.07   &    0.61 &    0.52 &    0.06 &    2.86 \\
M0204  &        3.72   &    0.28 &    0.24 &    0.09 &    2.50 \\
M0209  &        2.48   &    0.42 &    0.38 &    0.21 &    1.18 \\
M0215  &        3.16   &    0.31 &    0.25 &    0.16 &    1.04 \\
M0224  &        3.22   &    0.38 &    0.22 &    0.16 &    1.10 \\
M0227  &        4.53   &    0.37 &    0.28 &    0.27 &    1.42 \\
M0259  &        3.05   &    0.48 &    0.33 &    0.31 &    0.82 \\
M0290  &        2.17   &    0.40 &    0.35 &    0.40 &    0.57 \\
M0300  &        3.06   &    0.55 &    0.46 &    0.29 &    1.28 \\
M0329  &        2.96   &    0.37 &    0.32 &    0.07 &    1.24 \\
M0380  &        2.82   &    0.56 &    0.33 &    0.20 &    0.96 \\
M0408  &        2.59   &    0.49 &    0.46 &    0.38 &    0.63 \\
M0443  &        1.94   &    0.42 &    0.28 &    0.15 &    0.80 \\
M0549  &        3.39   &    0.44 &    0.23 &    0.17 &    0.98 \\
M0616  &        3.00   &    0.41 &    0.38 &    0.09 &    5.75 \\
M0664  &        2.16   &    0.36 &    0.24 &    0.11 &    3.25 \\
M0721  &        2.02   &    0.69 &    0.50 &    0.29 &    1.93 \\
M0763  &        3.32   &    0.41 &    0.41 &    0.44 &    1.05 \\
M0858  &        2.08   &    0.80 &    0.53 &    0.17 &    0.24 \\
M0908  &        2.09   &    0.61 &    0.45 &    0.31 &    0.44 \\
M0948  &        3.69   &    0.22 &    0.26 &    0.08 &    4.25 \\
M0959  &        2.29   &    0.29 &    0.35 &    0.12 &    5.11 \\
M1017  &        1.70   &    0.49 &    0.43 &    0.14 &    1.63 \\
M1061  &        2.55   &    0.51 &    0.49 &    0.26 &    8.60 \\
M1071  &        1.85   &    0.25 &    0.25 &    0.16 &    1.63 \\
M1091  &        1.58   &    0.50 &    0.47 &    0.12 &    9.13 \\
M1167  &        1.87   &    0.53 &    0.46 &    0.18 &    1.95 \\
M1192  &        2.29   &    0.59 &    0.44 &    0.16 &    0.88 \\
M1196  &        2.28   &    0.62 &    0.55 &    0.52 &    1.07 \\
M1306  &        1.47   &    0.65 &    0.43 &    0.28 &    0.51 \\
M1646  &        1.95   &    0.78 &    0.54 &    0.16 &    4.47 \\
M1859  &        1.60   &    0.42 &    0.38 &    0.22 &    5.14 \\
M2283  &        1.43   &    0.64 &    0.53 &    0.19 &    6.77 \\
M2665  &        1.57   &    0.37 &    0.31 &    0.16 &    3.16 \\
M3431  &        0.98   &    0.36 &    0.33 &    0.16 &    4.95 \\
M3852  &        1.49   &    0.56 &    0.51 &    0.63 &    0.97 \\
M4323  &        1.35   &    0.43 &    0.41 &    0.23 &    2.26 \\
M5014  &        1.33   &    0.62 &    0.56 &    0.53 &    1.71 \\
M6782  &        1.30   &    0.71 &    0.64 &    0.57 &    3.63 \\
\hline
\end{tabular}
\label{fitting}
\end{center}
\end{table}

\begin{table*}
\begin{center}
  \caption{Stellar mass and circular velocity curve parameters for the
    simulated galaxy sample: designations of the model galaxies
    ($1${\it st} column), stellar mass within $10\%~r_{\rm vir}$
    ($2${\it nd} column), circular velocities and their slopes at
    radius $5\re$ ($3${\it rd} and $4${\it th} columns), circular
    velocities from dark halo components only and their corresponding
    slopes at $5\re$ ($5${\it th} and $6${\it th} columns), parameters
    of the dark matter halo fit of eq.~\ref{vceq} ($7${\it th} to
    $9${\it th} columns), and rms residuals of this fit ($10${\it th}
    column).}
\begin{tabular}{lrcrcrcrcrc}
\hline
Model & Mass*  & $v_{\rm circ,5\re}$  & Slope of &  $v_{\rm circ,5\re}^{\rm DM}$ &  Slope of &$v_0 $ & a & $r_c $ & RMS\\
      & ($10^{10}\Msun~{\rm h^{-1}}$) &$(\kms)$ & $v_{\rm circ,5\re}$ & $(\kms)$         &  $v_{\rm circ,5\re}^{\rm DM}$ &$(\kms)$ &  & $(\kpc~{\rm h^{-1}})$ & $\kms$ \\
\hline
M0094  &     34.49 &    388.80 &      0.02 &    294.32 &      0.24 &    296.00 &      0.20 &      1.31 &      1.55 \\
M0125  &     31.22 &    375.83 &      0.03 &    298.30 &      0.23 &    297.60 &      0.19 &      1.42 &      1.57 \\
M0175  &     26.49 &    372.44 &     -0.03 &    299.45 &      0.14 &    296.18 &      0.16 &      1.45 &      3.71 \\
M0190  &     22.67 &    320.43 &     -0.08 &    241.16 &      0.09 &    243.62 &      0.13 &      0.92 &      1.94 \\
M0204  &     19.34 &    303.63 &      0.03 &    229.31 &      0.27 &    230.79 &      0.22 &      0.79 &      2.08 \\
M0209  &     14.37 &    309.16 &     -0.09 &    232.47 &      0.13 &    236.07 &      0.06 &      1.75 &      1.64 \\
M0215  &     19.90 &    337.30 &      0.01 &    259.32 &      0.22 &    261.00 &      0.14 &      1.33 &      1.61 \\
M0224  &     17.89 &    304.75 &      0.02 &    227.35 &      0.23 &    230.15 &      0.14 &      0.68 &      2.94 \\
M0227  &     22.23 &    317.13 &      0.00 &    244.73 &      0.18 &    247.19 &      0.18 &      0.94 &      2.04 \\
M0259  &     14.28 &    297.33 &     -0.06 &    227.16 &      0.13 &    226.97 &      0.12 &      1.25 &      1.58 \\
M0290  &     15.87 &    319.75 &     -0.10 &    225.31 &      0.12 &    229.24 &      0.11 &      1.20 &      1.96 \\
M0300  &     13.43 &    277.36 &     -0.10 &    204.18 &      0.11 &    207.10 &      0.09 &      0.98 &      1.69 \\
M0329  &     15.36 &    309.59 &     -0.05 &    235.69 &      0.18 &    237.27 &      0.11 &      1.55 &      0.82 \\
M0380  &     12.29 &    309.77 &     -0.06 &    249.41 &      0.10 &    248.51 &      0.09 &      1.46 &      2.51 \\
M0408  &     12.75 &    297.20 &     -0.13 &    217.43 &      0.06 &    219.60 &      0.06 &      1.15 &      2.00 \\
M0443  &     16.62 &    345.03 &     -0.09 &    235.38 &      0.20 &    237.56 &      0.09 &      1.13 &      1.24 \\
M0549  &      8.38 &    237.34 &      0.04 &    190.32 &      0.19 &    189.23 &      0.13 &      0.77 &      2.06 \\
M0616  &      9.38 &    259.79 &     -0.09 &    203.62 &      0.09 &    206.24 &      0.11 &      0.92 &      2.38 \\
M0664  &      7.48 &    246.16 &     -0.09 &    184.48 &      0.14 &    187.15 &      0.07 &      0.98 &      2.99 \\
M0721  &      9.63 &    276.12 &     -0.24 &    193.45 &     -0.05 &    198.68 &     -0.02 &      1.55 &      2.76 \\
M0763  &      9.85 &    238.93 &     -0.09 &    179.72 &      0.04 &    183.14 &      0.01 &      1.29 &      3.73 \\
M0858  &     10.27 &    264.88 &     -0.23 &    171.90 &      0.02 &    176.39 &     -0.01 &      1.46 &      2.42 \\
M0908  &      9.67 &    264.42 &     -0.21 &    176.03 &      0.02 &    178.18 &      0.01 &      1.21 &      1.44 \\
M0948  &      6.64 &    237.28 &     -0.02 &    198.55 &      0.11 &    197.98 &      0.10 &      1.41 &      2.31 \\
M0959  &      6.05 &    222.72 &     -0.06 &    167.71 &      0.15 &    171.47 &      0.02 &      1.52 &      2.95 \\
M1017  &      6.39 &    254.50 &     -0.18 &    182.86 &     -0.01 &    185.65 &      0.01 &      1.14 &      0.97 \\
M1061  &      5.18 &    206.25 &     -0.15 &    157.89 &      0.02 &    159.55 &      0.04 &      1.24 &      1.01 \\
M1071  &      7.79 &    240.60 &     -0.17 &    157.46 &      0.06 &    159.63 &      0.07 &      0.91 &      1.88 \\
M1091  &      7.53 &    243.13 &     -0.21 &    152.81 &      0.01 &    155.36 &     -0.02 &      1.03 &      1.10 \\
M1167  &      7.37 &    249.04 &     -0.17 &    172.20 &      0.06 &    173.63 &      0.04 &      1.12 &      1.09 \\
M1192  &      4.36 &    189.01 &     -0.07 &    141.12 &      0.12 &    144.20 &      0.06 &      1.05 &      3.17 \\
M1196  &      7.73 &    257.84 &     -0.19 &    188.89 &     -0.01 &    189.65 &      0.00 &      1.46 &      1.50 \\
M1306  &      6.51 &    245.04 &     -0.21 &    156.97 &      0.01 &    160.81 &      0.01 &      1.09 &      2.18 \\
M1646  &      5.38 &    224.42 &     -0.20 &    161.42 &      0.01 &    164.57 &     -0.02 &      1.30 &      1.57 \\
M1859  &      4.51 &    211.94 &     -0.13 &    149.00 &      0.08 &    152.64 &      0.01 &      1.02 &      1.70 \\
M2283  &      3.39 &    187.63 &     -0.17 &    126.38 &      0.05 &    128.99 &     -0.05 &      1.23 &      1.92 \\
M2665  &      3.17 &    185.07 &     -0.22 &    131.70 &     -0.03 &    135.47 &     -0.02 &      1.33 &      1.97 \\
M3431  &      1.87 &    175.82 &     -0.25 &    119.96 &     -0.02 &    125.09 &     -0.10 &      1.34 &      1.86 \\
M3852  &      2.64 &    173.24 &     -0.23 &    123.50 &     -0.09 &    127.22 &     -0.15 &      1.42 &      1.60 \\
M4323  &      2.38 &    169.72 &     -0.27 &    114.39 &     -0.09 &    116.01 &     -0.08 &      1.05 &      1.09 \\
M5014  &      2.26 &    163.71 &     -0.30 &    109.18 &     -0.12 &    116.20 &     -0.15 &      1.73 &      3.77 \\
M6782  &      1.95 &    158.66 &     -0.28 &    112.41 &     -0.10 &    116.33 &     -0.16 &      1.42 &      1.60 \\
\hline
\end{tabular}\label{tabvc}
\end{center}
\end{table*}

\bibliographystyle{mn2e}
\bibliography{cosgal}

\begin{thebibliography}{}

\bibitem[\protect\citeauthoryear{{Abadi}, {Navarro} \& {Steinmetz}}{{Abadi}
  et~al.}{2006}]{Abadi_etal2006}
{Abadi} M.~G.,  {Navarro} J.~F.,    {Steinmetz} M.,  2006, \mnras, 365, 747

\bibitem[\protect\citeauthoryear{{Allgood}, {Flores}, {Primack}, {Kravtsov},
  {Wechsler}, {Faltenbacher} \& {Bullock}}{{Allgood}
  et~al.}{2006}]{Allgood_etal2006}
{Allgood} B.,  {Flores} R.~A.,  {Primack} J.~R.,  {Kravtsov} A.~V.,  {Wechsler}
  R.~H.,  {Faltenbacher} A.,    {Bullock} J.~S.,  2006, \mnras, 367, 1781

\bibitem[\protect\citeauthoryear{{Auger}, {Treu}, {Bolton}, {Gavazzi},
  {Koopmans}, {Marshall}, {Moustakas} \& {Burles}}{{Auger}
  et~al.}{2010}]{Auger_etal2010}
{Auger} M.~W.,  {Treu} T.,  {Bolton} A.~S.,  {Gavazzi} R.,  {Koopmans}
  L.~V.~E.,  {Marshall} P.~J.,  {Moustakas} L.~A.,    {Burles} S.,  2010, \apj,
  724, 511

\bibitem[\protect\citeauthoryear{{Bailin}, {Kawata}, {Gibson}, {Steinmetz},
  {Navarro}, {Brook}, {Gill}, {Ibata}, {Knebe}, {Lewis} \& {Okamoto}}{{Bailin}
  et~al.}{2005}]{Bailin_etal2005}
{Bailin} J.,  {Kawata} D.,  {Gibson} B.~K.,  {Steinmetz} M.,  {Navarro} J.~F.,
  {Brook} C.~B.,  {Gill} S.~P.~D.,  {Ibata} R.~A.,  {Knebe} A.,  {Lewis} G.~F.,
     {Okamoto} T.,  2005, \apjl, 627, L17

\bibitem[\protect\citeauthoryear{{Barnab{\`e}}, {Czoske}, {Koopmans}, {Treu} \&
  {Bolton}}{{Barnab{\`e}} et~al.}{2011}]{Barnabe_etal2011}
{Barnab{\`e}} M.,  {Czoske} O.,  {Koopmans} L.~V.~E.,  {Treu} T.,    {Bolton}
  A.~S.,  2011, \mnras, 415, 2215

\bibitem[\protect\citeauthoryear{{Behroozi}, {Conroy} \& {Wechsler}}{{Behroozi}
  et~al.}{2010}]{Behroozi_etal2010}
{Behroozi} P.~S.,  {Conroy} C.,    {Wechsler} R.~H.,  2010, \apj, 717, 379

\bibitem[\protect\citeauthoryear{{Bender}, {Saglia} \& {Gerhard}}{{Bender}
  et~al.}{1994}]{Bender_etal1994}
{Bender} R.,  {Saglia} R.~P.,    {Gerhard} O.~E.,  1994, \mnras, 269, 785

\bibitem[\protect\citeauthoryear{{Bender}, {Surma}, {Doebereiner},
  {Moellenhoff} \& {Madejsky}}{{Bender} et~al.}{1989}]{Bender_etal1989}
{Bender} R.,  {Surma} P.,  {Doebereiner} S.,  {Moellenhoff} C.,    {Madejsky}
  R.,  1989, \aap, 217, 35

\bibitem[\protect\citeauthoryear{{Bernardi}, {Sheth}, {Annis}, {Burles},
  {Finkbeiner}, {Lupton}, {Schlegel}, {SubbaRao}, {Bahcall}, {Blakeslee},
  {Brinkmann}, {Castander}, {Connolly} \& {et al.}}{{Bernardi}
  et~al.}{2003}]{Bernardi_etal2003}
{Bernardi} M.,  {Sheth} R.~K.,  {Annis} J.,  {Burles} S.,  {Finkbeiner} D.~P.,
  {Lupton} R.~H.,  {Schlegel} D.~J.,  {SubbaRao} M.,  {Bahcall} N.~A.,
  {Blakeslee} J.~P.,  {Brinkmann} J.,  {Castander} F.~J.,  {Connolly} A.~J.,
  {et al.} 2003, \aj, 125, 1882

\bibitem[\protect\citeauthoryear{{Bett}, {Eke}, {Frenk}, {Jenkins} \&
  {Okamoto}}{{Bett} et~al.}{2010}]{Bett_etal2010}
{Bett} P.,  {Eke} V.,  {Frenk} C.~S.,  {Jenkins} A.,    {Okamoto} T.,  2010,
  \mnras, 404, 1137

\bibitem[\protect\citeauthoryear{{Brook}, {Stinson}, {Gibson}, {Ro{\v s}kar},
  {Wadsley} \& {Quinn}}{{Brook} et~al.}{2012}]{Brook_etal2012}
{Brook} C.~B.,  {Stinson} G.,  {Gibson} B.~K.,  {Ro{\v s}kar} R.,  {Wadsley}
  J.,    {Quinn} T.,  2012, \mnras, 419, 771

\bibitem[\protect\citeauthoryear{{Capaccioli}}{{Capaccioli}}{1989}]{Capaccioli%
1989}
{Capaccioli} M.,  1989, in {H.~G.~Corwin Jr.~\& L.~Bottinelli} ed., World of
  Galaxies (Le Monde des Galaxies) {Photometry of early-type galaxies and the R
  exp 1/4 law}.
pp 208--227

\bibitem[\protect\citeauthoryear{{Cappellari}, {Bacon}, {Bureau}, {Damen},
  {Davies}, {de Zeeuw}, {Emsellem}, {Falc{\'o}n-Barroso}, {Krajnovi{\'c}},
  {Kuntschner}, {McDermid}, {Peletier}, {Sarzi}, {van den Bosch} \& {van de
  Ven}}{{Cappellari} et~al.}{2006}]{Cappellari_etal2006}
{Cappellari} M.,  {Bacon} R.,  {Bureau} M.,  {Damen} M.~C.,  {Davies} R.~L.,
  {de Zeeuw} P.~T.,  {Emsellem} E.,  {Falc{\'o}n-Barroso} J.,  {Krajnovi{\'c}}
  D.,  {Kuntschner} H.,  {McDermid} R.~M.,  {Peletier} R.~F.,  {Sarzi} M.,
  {van den Bosch} R.~C.~E.,    {van de Ven} G.,  2006, \mnras, 366, 1126

\bibitem[\protect\citeauthoryear{{Cappellari}, {Emsellem}, {Bacon}, {Bureau},
  {Davies}, {de Zeeuw}, {Falc{\'o}n-Barroso}, {Krajnovi{\'c}}, {Kuntschner},
  {McDermid}, {Peletier}, {Sarzi}, {van den Bosch} \& {van de
  Ven}}{{Cappellari} et~al.}{2007}]{Cappellari_etal2007}
{Cappellari} M.,  {Emsellem} E.,  {Bacon} R.,  {Bureau} M.,  {Davies} R.~L.,
  {de Zeeuw} P.~T.,  {Falc{\'o}n-Barroso} J.,  {Krajnovi{\'c}} D.,
  {Kuntschner} H.,  {McDermid} R.~M.,  {Peletier} R.~F.,  {Sarzi} M.,  {van den
  Bosch} R.~C.~E.,    {van de Ven} G.,  2007, \mnras, 379, 418

\bibitem[\protect\citeauthoryear{{Cassata}, {Giavalisco}, {Guo}, {Renzini},
  {Ferguson}, {Koekemoer}, {Salimbeni}, {Scarlata}, {Grogin}, {Conselice},
  {Dahlen}, {Lotz}, {Dickinson} \& {Lin}}{{Cassata}
  et~al.}{2011}]{Cassata_etal2011}
{Cassata} P.,  {Giavalisco} M.,  {Guo} Y.,  {Renzini} A.,  {Ferguson} H.,
  {Koekemoer} A.~M.,  {Salimbeni} S.,  {Scarlata} C.,  {Grogin} N.~A.,
  {Conselice} C.~J.,  {Dahlen} T.,  {Lotz} J.~M.,  {Dickinson} M.,    {Lin} L.,
   2011, \apj, 743, 96

\bibitem[\protect\citeauthoryear{{Churazov}, {Forman}, {Vikhlinin}, {Tremaine},
  {Gerhard} \& {Jones}}{{Churazov} et~al.}{2008}]{Churazov_etal2008}
{Churazov} E.,  {Forman} W.,  {Vikhlinin} A.,  {Tremaine} S.,  {Gerhard} O.,
  {Jones} C.,  2008, \mnras, 388, 1062

\bibitem[\protect\citeauthoryear{{Churazov}, {Tremaine}, {Forman}, {Gerhard},
  {Das}, {Vikhlinin}, {Jones}, {B{\"o}hringer} \& {Gebhardt}}{{Churazov}
  et~al.}{2010}]{Churazov_etal2010}
{Churazov} E.,  {Tremaine} S.,  {Forman} W.,  {Gerhard} O.,  {Das} P.,
  {Vikhlinin} A.,  {Jones} C.,  {B{\"o}hringer} H.,    {Gebhardt} K.,  2010,
  \mnras, 404, 1165

\bibitem[\protect\citeauthoryear{{Coccato}, {Arnaboldi} \& {Gerhard}}{{Coccato}
  et~al.}{2013}]{Coccato_etal2012}
{Coccato} L.,  {Arnaboldi} M.,    {Gerhard} O.,  2013, \mnras

\bibitem[\protect\citeauthoryear{{Coccato}, {Gerhard}, {Arnaboldi}, {Das},
  {Douglas}, {Kuijken}, {Merrifield}, {Napolitano}, {Noordermeer},
  {Romanowsky}, {Capaccioli}, {Cortesi}, {de Lorenzi} \& {Freeman}}{{Coccato}
  et~al.}{2009}]{Coccato_etal2009}
{Coccato} L.,  {Gerhard} O.,  {Arnaboldi} M.,  {Das} P.,  {Douglas} N.~G.,
  {Kuijken} K.,  {Merrifield} M.~R.,  {Napolitano} N.~R.,  {Noordermeer} E.,
  {Romanowsky} A.~J.,  {Capaccioli} M.,  {Cortesi} A.,  {de Lorenzi} F.,
  {Freeman} K.~C.,  2009, \mnras, 394, 1249

\bibitem[\protect\citeauthoryear{{Croton}, {Springel}, {White}, {De Lucia},
  {Frenk}, {Gao}, {Jenkins}, {Kauffmann}, {Navarro} \& {Yoshida}}{{Croton}
  et~al.}{2006}]{Croton_etal2006}
{Croton} D.~J.,  {Springel} V.,  {White} S.~D.~M.,  {De Lucia} G.,  {Frenk}
  C.~S.,  {Gao} L.,  {Jenkins} A.,  {Kauffmann} G.,  {Navarro} J.~F.,
  {Yoshida} N.,  2006, \mnras, 365, 11

\bibitem[\protect\citeauthoryear{{Daddi}, {Renzini}, {Pirzkal}, {Cimatti},
  {Malhotra}, {Stiavelli}, {Xu}, {Pasquali}, {Rhoads}, {Brusa}, {di Serego
  Alighieri}, {Ferguson}, {Koekemoer}, {Moustakas}, {Panagia} \&
  {Windhorst}}{{Daddi} et~al.}{2005}]{Daddi_etal2005}
{Daddi} E.,  {Renzini} A.,  {Pirzkal} N.,  {Cimatti} A.,  {Malhotra} S.,
  {Stiavelli} M.,  {Xu} C.,  {Pasquali} A.,  {Rhoads} J.~E.,  {Brusa} M.,  {di
  Serego Alighieri} S.,  {Ferguson} H.~C.,  {Koekemoer} A.~M.,  {Moustakas}
  L.~A.,  {Panagia} N.,    {Windhorst} R.~A.,  2005, \apj, 626, 680

\bibitem[\protect\citeauthoryear{{Dalla Vecchia} \& {Schaye}}{{Dalla Vecchia}
  \& {Schaye}}{2012}]{DallaVecchia+Schaye_2012}
{Dalla Vecchia} C.,  {Schaye} J.,  2012, ArXiv e-prints

\bibitem[\protect\citeauthoryear{{de Lorenzi}, {Debattista}, {Gerhard} \&
  {Sambhus}}{{de Lorenzi} et~al.}{2007}]{nmagic}
{de Lorenzi} F.,  {Debattista} V.~P.,  {Gerhard} O.,    {Sambhus} N.,  2007,
  \mnras, 376, 71

\bibitem[\protect\citeauthoryear{{de Lorenzi}, {Gerhard}, {Coccato},
  {Arnaboldi}, {Capaccioli}, {Douglas}, {Freeman}, {Kuijken}, {Merrifield},
  {Napolitano}, {Noordermeer}, {Romanowsky} \& {Debattista}}{{de Lorenzi}
  et~al.}{2009}]{ngc3379}
{de Lorenzi} F.,  {Gerhard} O.,  {Coccato} L.,  {Arnaboldi} M.,  {Capaccioli}
  M.,  {Douglas} N.~G.,  {Freeman} K.~C.,  {Kuijken} K.,  {Merrifield} M.~R.,
  {Napolitano} N.~R.,  {Noordermeer} E.,  {Romanowsky} A.~J.,    {Debattista}
  V.~P.,  2009, \mnras, 395, 76

\bibitem[\protect\citeauthoryear{{de Lorenzi}, {Gerhard}, {Saglia}, {Sambhus},
  {Debattista}, {Pannella} \& {M{\'e}ndez}}{{de Lorenzi}
  et~al.}{2008}]{ngc4697}
{de Lorenzi} F.,  {Gerhard} O.,  {Saglia} R.~P.,  {Sambhus} N.,  {Debattista}
  V.~P.,  {Pannella} M.,    {M{\'e}ndez} R.~H.,  2008, \mnras, 385, 1729

\bibitem[\protect\citeauthoryear{{Deason}, {Belokurov}, {Evans} \&
  {McCarthy}}{{Deason} et~al.}{2012}]{Deason_etal2012}
{Deason} A.~J.,  {Belokurov} V.,  {Evans} N.~W.,    {McCarthy} I.~G.,  2012,
  \apj, 748, 2

\bibitem[\protect\citeauthoryear{{Dekel} \& {Silk}}{{Dekel} \&
  {Silk}}{1986}]{Dekel+Silk_1986}
{Dekel} A.,  {Silk} J.,  1986, \apj, 303, 39

\bibitem[\protect\citeauthoryear{{Dekel}, {Stoehr}, {Mamon}, {Cox}, {Novak} \&
  {Primack}}{{Dekel} et~al.}{2005}]{Dekel_etal2005}
{Dekel} A.,  {Stoehr} F.,  {Mamon} G.~A.,  {Cox} T.~J.,  {Novak} G.~S.,
  {Primack} J.~R.,  2005, \nat, 437, 707

\bibitem[\protect\citeauthoryear{{Di Matteo}, {Colberg}, {Springel},
  {Hernquist} \& {Sijacki}}{{Di Matteo} et~al.}{2008}]{DiMatteo_etal2008}
{Di Matteo} T.,  {Colberg} J.,  {Springel} V.,  {Hernquist} L.,    {Sijacki}
  D.,  2008, \apj, 676, 33

\bibitem[\protect\citeauthoryear{{Duc}, {Cuillandre}, {Serra}, {Michel-Dansac},
  {Ferriere}, {Alatalo}, {Blitz}, {Bois}, {Bournaud}, {Bureau}, {Cappellari},
  {Davies} \& {et al.}}{{Duc} et~al.}{2011}]{Duc_etal2011}
{Duc} P.-A.,  {Cuillandre} J.-C.,  {Serra} P.,  {Michel-Dansac} L.,  {Ferriere}
  E.,  {Alatalo} K.,  {Blitz} L.,  {Bois} M.,  {Bournaud} F.,  {Bureau} M.,
  {Cappellari} M.,  {Davies} R.~L.,    {et al.} 2011, \mnras, 417, 863

\bibitem[\protect\citeauthoryear{{Emsellem}, {Cappellari}, {Krajnovi{\'c}},
  {Alatalo}, {Blitz}, {Bois}, {Bournaud}, {Bureau} \& {et al.}}{{Emsellem}
  et~al.}{2011}]{Emsellem_etal2011}
{Emsellem} E.,  {Cappellari} M.,  {Krajnovi{\'c}} D.,  {Alatalo} K.,  {Blitz}
  L.,  {Bois} M.,  {Bournaud} F.,  {Bureau} M.,    {et al.} 2011, \mnras, 414,
  888

\bibitem[\protect\citeauthoryear{{Emsellem}, {Cappellari}, {Krajnovi{\'c}},
  {van de Ven}, {Bacon}, {Bureau}, {Davies}, {de Zeeuw}, {Falc{\'o}n-Barroso},
  {Kuntschner}, {McDermid}, {Peletier} \& {Sarzi}}{{Emsellem}
  et~al.}{2007}]{Emsellem_etal2007}
{Emsellem} E.,  {Cappellari} M.,  {Krajnovi{\'c}} D.,  {van de Ven} G.,
  {Bacon} R.,  {Bureau} M.,  {Davies} R.~L.,  {de Zeeuw} P.~T.,
  {Falc{\'o}n-Barroso} J.,  {Kuntschner} H.,  {McDermid} R.,  {Peletier} R.~F.,
     {Sarzi} M.,  2007, \mnras, 379, 401

\bibitem[\protect\citeauthoryear{{Feldmann}, {Carollo}, {Mayer}, {Renzini},
  {Lake}, {Quinn}, {Stinson} \& {Yepes}}{{Feldmann}
  et~al.}{2010}]{Feldmann_etal2010}
{Feldmann} R.,  {Carollo} C.~M.,  {Mayer} L.,  {Renzini} A.,  {Lake} G.,
  {Quinn} T.,  {Stinson} G.~S.,    {Yepes} G.,  2010, \apj, 709, 218

\bibitem[\protect\citeauthoryear{{Fontana}, {Salimbeni}, {Grazian},
  {Giallongo}, {Pentericci}, {Nonino}, {Fontanot}, {Menci}, {Monaco},
  {Cristiani}, {Vanzella}, {de Santis} \& {Gallozzi}}{{Fontana}
  et~al.}{2006}]{Fontana_etal2006}
{Fontana} A.,  {Salimbeni} S.,  {Grazian} A.,  {Giallongo} E.,  {Pentericci}
  L.,  {Nonino} M.,  {Fontanot} F.,  {Menci} N.,  {Monaco} P.,  {Cristiani} S.,
   {Vanzella} E.,  {de Santis} C.,    {Gallozzi} S.,  2006, \aap, 459, 745

\bibitem[\protect\citeauthoryear{{Gerhard}, {Kronawitter}, {Saglia} \&
  {Bender}}{{Gerhard} et~al.}{2001}]{Gerhard_etal2001}
{Gerhard} O.,  {Kronawitter} A.,  {Saglia} R.~P.,    {Bender} R.,  2001, \aj,
  121, 1936

\bibitem[\protect\citeauthoryear{{Governato}, {Brook}, {Mayer}, {Brooks},
  {Rhee}, {Wadsley}, {Jonsson}, {Willman}, {Stinson}, {Quinn} \&
  {Madau}}{{Governato} et~al.}{2010}]{Governato_etal2010}
{Governato} F.,  {Brook} C.,  {Mayer} L.,  {Brooks} A.,  {Rhee} G.,  {Wadsley}
  J.,  {Jonsson} P.,  {Willman} B.,  {Stinson} G.,  {Quinn} T.,    {Madau} P.,
  2010, \nat, 463, 203

\bibitem[\protect\citeauthoryear{{Graham}, {Erwin}, {Trujillo} \& {Asensio
  Ramos}}{{Graham} et~al.}{2003}]{Graham_etal2003}
{Graham} A.~W.,  {Erwin} P.,  {Trujillo} I.,    {Asensio Ramos} A.,  2003, \aj,
  125, 2951

\bibitem[\protect\citeauthoryear{{Guedes}, {Callegari}, {Madau} \&
  {Mayer}}{{Guedes} et~al.}{2011}]{Guedes_etal2011}
{Guedes} J.,  {Callegari} S.,  {Madau} P.,    {Mayer} L.,  2011, \apj, 742, 76

\bibitem[\protect\citeauthoryear{{Guo}, {White}, {Li} \&
  {Boylan-Kolchin}}{{Guo} et~al.}{2010}]{Guo_etal2010}
{Guo} Q.,  {White} S.,  {Li} C.,    {Boylan-Kolchin} M.,  2010, \mnras, 404,
  1111

\bibitem[\protect\citeauthoryear{{Hahn}, {Teyssier} \& {Carollo}}{{Hahn}
  et~al.}{2010}]{Hahn_etal2010}
{Hahn} O.,  {Teyssier} R.,    {Carollo} C.~M.,  2010, \mnras, 405, 274

\bibitem[\protect\citeauthoryear{{Hilz}, {Naab} \& {Ostriker}}{{Hilz}
  et~al.}{2013}]{Hilz_etal2013}
{Hilz} M.,  {Naab} T.,    {Ostriker} J.~P.,  2013, \mnras, 429, 2924

\bibitem[\protect\citeauthoryear{{Hilz}, {Naab}, {Ostriker}, {Thomas},
  {Burkert} \& {Jesseit}}{{Hilz} et~al.}{2012}]{Hilz_etal2012}
{Hilz} M.,  {Naab} T.,  {Ostriker} J.~P.,  {Thomas} J.,  {Burkert} A.,
  {Jesseit} R.,  2012, \mnras, 425, 3119

\bibitem[\protect\citeauthoryear{{Hoekstra}, {Yee} \& {Gladders}}{{Hoekstra}
  et~al.}{2004}]{Hoekstra_etal2004}
{Hoekstra} H.,  {Yee} H.~K.~C.,    {Gladders} M.~D.,  2004, \apj, 606, 67

\bibitem[\protect\citeauthoryear{{Hopkins}, {Bundy}, {Hernquist}, {Wuyts} \&
  {Cox}}{{Hopkins} et~al.}{2010}]{Hopkins_etal2010}
{Hopkins} P.~F.,  {Bundy} K.,  {Hernquist} L.,  {Wuyts} S.,    {Cox} T.~J.,
  2010, \mnras, 401, 1099

\bibitem[\protect\citeauthoryear{{Ilbert}, {Salvato}, {Le Floc'h}, {Aussel},
  {Capak}, {McCracken}, {Mobasher} \& {et al}}{{Ilbert}
  et~al.}{2010}]{Ilbert_etal2010}
{Ilbert} O.,  {Salvato} M.,  {Le Floc'h} E.,  {Aussel} H.,  {Capak} P.,
  {McCracken} H.~J.,  {Mobasher} B.,    {et al} 2010, \apj, 709, 644

\bibitem[\protect\citeauthoryear{{Jesseit}, {Cappellari}, {Naab}, {Emsellem} \&
  {Burkert}}{{Jesseit} et~al.}{2009}]{Jesseit_etal2009}
{Jesseit} R.,  {Cappellari} M.,  {Naab} T.,  {Emsellem} E.,    {Burkert} A.,
  2009, \mnras, 397, 1202

\bibitem[\protect\citeauthoryear{{Jing} \& {Suto}}{{Jing} \&
  {Suto}}{2002}]{Jing_suto2002}
{Jing} Y.~P.,  {Suto} Y.,  2002, \apj, 574, 538

\bibitem[\protect\citeauthoryear{{Johansson}, {Naab} \& {Ostriker}}{{Johansson}
  et~al.}{2009}]{Johansson_etal2009}
{Johansson} P.~H.,  {Naab} T.,    {Ostriker} J.~P.,  2009, \apjl, 697, L38

\bibitem[\protect\citeauthoryear{{Johansson}, {Naab} \& {Ostriker}}{{Johansson}
  et~al.}{2012}]{Johansson_etal2012}
{Johansson} P.~H.,  {Naab} T.,    {Ostriker} J.~P.,  2012, \apj, 754, 115

\bibitem[\protect\citeauthoryear{{Kazantzidis}, {Kravtsov}, {Zentner},
  {Allgood}, {Nagai} \& {Moore}}{{Kazantzidis}
  et~al.}{2004}]{Kazantzidis_etal2004}
{Kazantzidis} S.,  {Kravtsov} A.~V.,  {Zentner} A.~R.,  {Allgood} B.,  {Nagai}
  D.,    {Moore} B.,  2004, \apjl, 611, L73

\bibitem[\protect\citeauthoryear{{Koopmans}, {Treu}, {Bolton}, {Burles} \&
  {Moustakas}}{{Koopmans} et~al.}{2006}]{Koopmans_etal2006}
{Koopmans} L.~V.~E.,  {Treu} T.,  {Bolton} A.~S.,  {Burles} S.,    {Moustakas}
  L.~A.,  2006, \apj, 649, 599

\bibitem[\protect\citeauthoryear{{Kormendy} \& {Bender}}{{Kormendy} \&
  {Bender}}{1996}]{Kormendy_Bender1996}
{Kormendy} J.,  {Bender} R.,  1996, \apjl, 464, L119+

\bibitem[\protect\citeauthoryear{{Kormendy}, {Fisher}, {Cornell} \&
  {Bender}}{{Kormendy} et~al.}{2009}]{Kormendy_etal2009}
{Kormendy} J.,  {Fisher} D.~B.,  {Cornell} M.~E.,    {Bender} R.,  2009, \apjs,
  182, 216

\bibitem[\protect\citeauthoryear{{Krajnovi{\'c}}, {Emsellem}, {Cappellari},
  {Alatalo}, {Blitz}, {Bois}, {Bournaud}, {Bureau} \& {et al.}}{{Krajnovi{\'c}}
  et~al.}{2011}]{Krajnovic_etal2011}
{Krajnovi{\'c}} D.,  {Emsellem} E.,  {Cappellari} M.,  {Alatalo} K.,  {Blitz}
  L.,  {Bois} M.,  {Bournaud} F.,  {Bureau} M.,    {et al.} 2011, \mnras, 414,
  2923

\bibitem[\protect\citeauthoryear{{Lyskova}, {Churazov}, {Zhuravleva}, {Naab},
  {Oser}, {Gerhard} \& {Wu}}{{Lyskova} et~al.}{2012}]{Lyskova_etal2012}
{Lyskova} N.,  {Churazov} E.,  {Zhuravleva} I.,  {Naab} T.,  {Oser} L.,
  {Gerhard} O.,    {Wu} X.,  2012, \mnras, 423, 1813

\bibitem[\protect\citeauthoryear{{Mandelbaum}, {Hirata}, {Broderick}, {Seljak}
  \& {Brinkmann}}{{Mandelbaum} et~al.}{2006}]{Mandelbaum_etal2006}
{Mandelbaum} R.,  {Hirata} C.~M.,  {Broderick} T.,  {Seljak} U.,    {Brinkmann}
  J.,  2006, \mnras, 370, 1008

\bibitem[\protect\citeauthoryear{{McCarthy}, {Schaye}, {Ponman}, {Bower},
  {Booth}, {Dalla Vecchia}, {Crain}, {Springel}, {Theuns} \&
  {Wiersma}}{{McCarthy} et~al.}{2010}]{McCarthy_etal2010}
{McCarthy} I.~G.,  {Schaye} J.,  {Ponman} T.~J.,  {Bower} R.~G.,  {Booth}
  C.~M.,  {Dalla Vecchia} C.,  {Crain} R.~A.,  {Springel} V.,  {Theuns} T.,
  {Wiersma} R.~P.~C.,  2010, \mnras, 406, 822

\bibitem[\protect\citeauthoryear{{McNeil-Moylan}, {Freeman}, {Arnaboldi} \&
  {Gerhard}}{{McNeil-Moylan} et~al.}{2012}]{McNeil-Moylan_etal2012}
{McNeil-Moylan} E.~K.,  {Freeman} K.~C.,  {Arnaboldi} M.,    {Gerhard} O.~E.,
  2012, \aap, 539, A11

\bibitem[\protect\citeauthoryear{{M{\'e}ndez}, {Riffeser}, {Kudritzki},
  {Matthias}, {Freeman}, {Arnaboldi}, {Capaccioli} \& {Gerhard}}{{M{\'e}ndez}
  et~al.}{2001}]{Mendez_etal2001}
{M{\'e}ndez} R.~H.,  {Riffeser} A.,  {Kudritzki} R.-P.,  {Matthias} M.,
  {Freeman} K.~C.,  {Arnaboldi} M.,  {Capaccioli} M.,    {Gerhard} O.~E.,
  2001, \apj, 563, 135

\bibitem[\protect\citeauthoryear{{Morganti}, {Gerhard}, {Coccato},
  {Martinez-Valpuesta} \& {Arnaboldi}}{{Morganti}
  et~al.}{2013}]{Morganti_etal2013}
{Morganti} L.,  {Gerhard} O.,  {Coccato} L.,  {Martinez-Valpuesta} I.,
  {Arnaboldi} M.,  2013, \mnras, 431, 3570

\bibitem[\protect\citeauthoryear{{Moster}, {Somerville}, {Maulbetsch}, {van den
  Bosch}, {Macci{\`o}}, {Naab} \& {Oser}}{{Moster}
  et~al.}{2010}]{Moster_etal2010}
{Moster} B.~P.,  {Somerville} R.~S.,  {Maulbetsch} C.,  {van den Bosch} F.~C.,
  {Macci{\`o}} A.~V.,  {Naab} T.,    {Oser} L.,  2010, \apj, 710, 903

\bibitem[\protect\citeauthoryear{{Murphy}, {Gebhardt} \& {Adams}}{{Murphy}
  et~al.}{2011}]{Murphy_etal2011}
{Murphy} J.~D.,  {Gebhardt} K.,    {Adams} J.~J.,  2011, \apj, 729, 129

\bibitem[\protect\citeauthoryear{{Naab}, {Johansson} \& {Ostriker}}{{Naab}
  et~al.}{2009}]{Naab_etal2009}
{Naab} T.,  {Johansson} P.~H.,    {Ostriker} J.~P.,  2009, \apjl, 699, L178

\bibitem[\protect\citeauthoryear{{Naab}, {Johansson}, {Ostriker} \&
  {Efstathiou}}{{Naab} et~al.}{2007}]{Naab_etal2007}
{Naab} T.,  {Johansson} P.~H.,  {Ostriker} J.~P.,    {Efstathiou} G.,  2007,
  \apj, 658, 710

\bibitem[\protect\citeauthoryear{{Naab}, {Oser}, {Emsellem}, {Cappellari},
  {Krajnovic}, {McDermid}, {Alatalo} \& {et al.}}{{Naab}
  et~al.}{2013}]{Naab_etal2013}
{Naab} T.,  {Oser} L.,  {Emsellem} E.,  {Cappellari} M.,  {Krajnovic} D.,
  {McDermid} R.~M.,  {Alatalo} K.,    {et al.} 2013, ArXiv e-prints

\bibitem[\protect\citeauthoryear{{Nagino} \& {Matsushita}}{{Nagino} \&
  {Matsushita}}{2009}]{Nagino_etal2009}
{Nagino} R.,  {Matsushita} K.,  2009, \aap, 501, 157

\bibitem[\protect\citeauthoryear{{Napolitano}, {Romanowsky}, {Coccato},
  {Capaccioli}, {Douglas}, {Noordermeer}, {Gerhard}, {Arnaboldi}, {de Lorenzi},
  {Kuijken}, {Merrifield}, {O'Sullivan}, {Cortesi}, {Das} \&
  {Freeman}}{{Napolitano} et~al.}{2009}]{Napolitano_etal2009}
{Napolitano} N.~R.,  {Romanowsky} A.~J.,  {Coccato} L.,  {Capaccioli} M.,
  {Douglas} N.~G.,  {Noordermeer} E.,  {Gerhard} O.,  {Arnaboldi} M.,  {de
  Lorenzi} F.,  {Kuijken} K.,  {Merrifield} M.~R.,  {O'Sullivan} E.,  {Cortesi}
  A.,  {Das} P.,    {Freeman} K.~C.,  2009, \mnras, 393, 329

\bibitem[\protect\citeauthoryear{{Navarro}, {Frenk} \& {White}}{{Navarro}
  et~al.}{1996}]{NFW1996}
{Navarro} J.~F.,  {Frenk} C.~S.,    {White} S.~D.~M.,  1996, \apj, 462, 563

\bibitem[\protect\citeauthoryear{{Navarro}, {Ludlow}, {Springel}, {Wang},
  {Vogelsberger}, {White}, {Jenkins}, {Frenk} \& {Helmi}}{{Navarro}
  et~al.}{2010}]{Navarro_etal2010}
{Navarro} J.~F.,  {Ludlow} A.,  {Springel} V.,  {Wang} J.,  {Vogelsberger} M.,
  {White} S.~D.~M.,  {Jenkins} A.,  {Frenk} C.~S.,    {Helmi} A.,  2010,
  \mnras, 402, 21

\bibitem[\protect\citeauthoryear{{Novak}, {Cox}, {Primack}, {Jonsson} \&
  {Dekel}}{{Novak} et~al.}{2006}]{Novak_etal2006}
{Novak} G.~S.,  {Cox} T.~J.,  {Primack} J.~R.,  {Jonsson} P.,    {Dekel} A.,
  2006, \apjl, 646, L9

\bibitem[\protect\citeauthoryear{{Oppenheimer} \& {Dav{\'e}}}{{Oppenheimer} \&
  {Dav{\'e}}}{2008}]{Oppenheimer+Dave_2008}
{Oppenheimer} B.~D.,  {Dav{\'e}} R.,  2008, \mnras, 387, 577

\bibitem[\protect\citeauthoryear{{Oser}, {Naab}, {Ostriker} \&
  {Johansson}}{{Oser} et~al.}{2012}]{Oser_etal2012}
{Oser} L.,  {Naab} T.,  {Ostriker} J.~P.,    {Johansson} P.~H.,  2012, \apj,
  744, 63

\bibitem[\protect\citeauthoryear{{Oser}, {Ostriker}, {Naab}, {Johansson} \&
  {Burkert}}{{Oser} et~al.}{2010}]{Oser_etal2010}
{Oser} L.,  {Ostriker} J.~P.,  {Naab} T.,  {Johansson} P.~H.,    {Burkert} A.,
  2010, \apj, 725, 2312

\bibitem[\protect\citeauthoryear{{Proctor}, {Forbes}, {Romanowsky}, {Brodie},
  {Strader}, {Spolaor}, {Mendel} \& {Spitler}}{{Proctor}
  et~al.}{2009}]{Proctor_etal2009}
{Proctor} R.~N.,  {Forbes} D.~A.,  {Romanowsky} A.~J.,  {Brodie} J.~P.,
  {Strader} J.,  {Spolaor} M.,  {Mendel} J.~T.,    {Spitler} L.,  2009, \mnras,
  398, 91

\bibitem[\protect\citeauthoryear{{Romanowsky} \& {Fall}}{{Romanowsky} \&
  {Fall}}{2012}]{Romanowsky+Fall_2012}
{Romanowsky} A.~J.,  {Fall} S.~M.,  2012, \apjs, 203, 17

\bibitem[\protect\citeauthoryear{{Schuberth}, {Richtler}, {Hilker}, {Dirsch},
  {Bassino}, {Romanowsky} \& {Infante}}{{Schuberth}
  et~al.}{2010}]{Schuberth_etal2010}
{Schuberth} Y.,  {Richtler} T.,  {Hilker} M.,  {Dirsch} B.,  {Bassino} L.~P.,
  {Romanowsky} A.~J.,    {Infante} L.,  2010, \aap, 513, A52

\bibitem[\protect\citeauthoryear{{S{\'e}rsic}}{{S{\'e}rsic}}{1963}]{Sersic1963}
{S{\'e}rsic} J.~L.,  1963, Boletin de la Asociacion Argentina de Astronomia La
  Plata Argentina, 6, 41

\bibitem[\protect\citeauthoryear{{Strader}, {Romanowsky}, {Brodie}, {Spitler},
  {Beasley}, {Arnold}, {Tamura}, {Sharples} \& {Arimoto}}{{Strader}
  et~al.}{2011}]{Strader_etal2011}
{Strader} J.,  {Romanowsky} A.~J.,  {Brodie} J.~P.,  {Spitler} L.~R.,
  {Beasley} M.~A.,  {Arnold} J.~A.,  {Tamura} N.,  {Sharples} R.~M.,
  {Arimoto} N.,  2011, \apjs, 197, 33

\bibitem[\protect\citeauthoryear{{Syer} \& {Tremaine}}{{Syer} \&
  {Tremaine}}{1996}]{Syer_tremaine1996}
{Syer} D.,  {Tremaine} S.,  1996, \mnras, 282, 223

\bibitem[\protect\citeauthoryear{{Teyssier}, {Moore}, {Martizzi}, {Dubois} \&
  {Mayer}}{{Teyssier} et~al.}{2011}]{Teyssier_etal2011}
{Teyssier} R.,  {Moore} B.,  {Martizzi} D.,  {Dubois} Y.,    {Mayer} L.,  2011,
  \mnras, 414, 195

\bibitem[\protect\citeauthoryear{{Trujillo}, {Conselice}, {Bundy}, {Cooper},
  {Eisenhardt} \& {Ellis}}{{Trujillo} et~al.}{2007}]{Trujillo_etal2007}
{Trujillo} I.,  {Conselice} C.~J.,  {Bundy} K.,  {Cooper} M.~C.,  {Eisenhardt}
  P.,    {Ellis} R.~S.,  2007, \mnras, 382, 109

\bibitem[\protect\citeauthoryear{{Trujillo}, {Erwin}, {Asensio Ramos} \&
  {Graham}}{{Trujillo} et~al.}{2004}]{Trujillo_etal2004}
{Trujillo} I.,  {Erwin} P.,  {Asensio Ramos} A.,    {Graham} A.~W.,  2004, \aj,
  127, 1917

\bibitem[\protect\citeauthoryear{{van Dokkum}}{{van
  Dokkum}}{2005}]{vanDokkum2005}
{van Dokkum} P.~G.,  2005, \aj, 130, 2647

\bibitem[\protect\citeauthoryear{{van Dokkum}, {Whitaker}, {Brammer}, {Franx},
  {Kriek} \& {et al.}}{{van Dokkum} et~al.}{2010}]{vanDokkum_etal2010}
{van Dokkum} P.~G.,  {Whitaker} K.~E.,  {Brammer} G.,  {Franx} M.,  {Kriek} M.,
     {et al.} 2010, \apj, 709, 1018

\bibitem[\protect\citeauthoryear{{van Uitert}, {Hoekstra}, {Schrabback},
  {Gilbank}, {Gladders} \& {Yee}}{{van Uitert}
  et~al.}{2012}]{vanUitert_etal2012}
{van Uitert} E.,  {Hoekstra} H.,  {Schrabback} T.,  {Gilbank} D.~G.,
  {Gladders} M.~D.,    {Yee} H.~K.~C.,  2012, ArXiv e-prints

\bibitem[\protect\citeauthoryear{{Weijmans}, {Cappellari}, {Bacon}, {de Zeeuw},
  {Emsellem}, {Falc{\'o}n-Barroso}, {Kuntschner}, {McDermid}, {van den Bosch}
  \& {van de Ven}}{{Weijmans} et~al.}{2009}]{Weijmans_etal2009}
{Weijmans} A.-M.,  {Cappellari} M.,  {Bacon} R.,  {de Zeeuw} P.~T.,  {Emsellem}
  E.,  {Falc{\'o}n-Barroso} J.,  {Kuntschner} H.,  {McDermid} R.~M.,  {van den
  Bosch} R.~C.~E.,    {van de Ven} G.,  2009, \mnras, 398, 561

\bibitem[\protect\citeauthoryear{{Yang}, {Mo}, {van den Bosch}, {Zhang} \&
  {Han}}{{Yang} et~al.}{2012}]{Yang_etal2012}
{Yang} X.,  {Mo} H.~J.,  {van den Bosch} F.~C.,  {Zhang} Y.,    {Han} J.,
  2012, \apj, 752, 41

\end{thebibliography}

\end{document}